\documentclass[a4paper,twocolumn,11pt]{quantumarticle}
\pdfoutput=1
\usepackage[utf8]{inputenc}
\usepackage[english]{babel}
\usepackage[T1]{fontenc}
\usepackage{tabularx}
\usepackage{array}
\usepackage[normalem]{ulem}
\usepackage[numbers,sort&compress]{natbib}

\usepackage{braket}
\usepackage{soul}
\usepackage{amsfonts, amssymb, amsmath, mathtools, bm}
\usepackage{siunitx}
\usepackage{booktabs}
\usepackage[dvipsnames]{xcolor}
\usepackage{xurl}
\usepackage{hyperref}
\usepackage{comment}
\usepackage{makecell}
\usepackage{multirow}
\usepackage{orcidlink}
\usepackage{graphicx}
\usepackage{svg}
\usepackage{physics}
\usepackage{listings}
\usepackage{color}
\usepackage[dvipsnames]{xcolor}
\usepackage{pagecolor}
\usepackage[capitalise]{cleveref}
\usepackage{tikz}
\usetikzlibrary{quantikz2}
\definecolor{ionq_orange}{HTML}{ff5000}
\hypersetup{colorlinks=true, linkcolor=Blue, citecolor=Blue, urlcolor=Blue}

\newcommand{\btheta}{\boldsymbol{ \theta}}

\newcolumntype{L}{>{\hsize=0.6\hsize}X}
\newcolumntype{W}{>{\hsize=1.1\hsize}X}

\begin{document}

\title{A Non-Variational Quantum Approach to the Job Shop Scheduling Problem}

\author{Miguel Angel Lopez-Ruiz}
\email[]{miguel.lopezruiz@ionq.co}
\affiliation{IonQ Inc., 4505 Campus Dr, College Park, MD 20740, USA}
\orcid{0000-0002-8152-5655}

\author{Emily L. Tucker}
\email{etucke3@clemson.edu}
\affiliation{Department of Industrial Engineering, 211 Fernow St, Clemson University, Clemson, SC 29634-0920, USA}
\orcid{0000-0003-1370-1642}

\author{Emma M. Arnold}
\affiliation{Department of Industrial Engineering, 211 Fernow St, Clemson University, Clemson, SC 29634-0920, USA}
\orcid{0009-0005-8393-323X}

\author{Evgeny Epifanovsky}
\affiliation{IonQ Inc., 4505 Campus Dr, College Park, MD 20740, USA}
\orcid{0000-0002-3379-1131}

\author{Ananth Kaushik}
\affiliation{IonQ Inc., 4505 Campus Dr, College Park, MD 20740, USA}
\orcid{0009-0009-2799-1194}

\author{Martin Roetteler}
\affiliation{IonQ Inc., 4505 Campus Dr, College Park, MD 20740, USA}
\orcid{0000-0003-0234-2496}

\begin{abstract}
\noindent
   Quantum heuristics offer a potential advantage for combinatorial optimization but are constrained by near-term hardware limitations. We introduce Iterative-QAOA, a variant of QAOA designed to mitigate these constraints. The algorithm combines a non-variational, shallow-depth circuit approach using fixed-parameter schedules with an iterative warm-starting process. We benchmark the algorithm on Just-in-Time Job Shop Scheduling Problem (JIT-JSSP) instances on IonQ Forte Generation QPUs, representing some of the largest such problems ever executed on quantum hardware. We compare the performance of the algorithm against both the Variational Quantum Imaginary Time Evolution (VarQITE) algorithm and the non-variational Linear Ramp (LR) QAOA algorithm. 
   We find that Iterative-QAOA robustly converges to find optimal solutions as well as high-quality, near-optimal solutions for all problem instances evaluated. We evaluate the algorithm on larger problem instances up to 97 qubits using tensor network simulations. The scaling behavior of the algorithm indicates potential for solving industrial-scale problems on fault-tolerant quantum computers.
\end{abstract}

\maketitle

\section{Introduction}

\noindent
The Job Shop Scheduling Problem (JSSP) is among the most widely studied combinatorial optimization problems \cite{Xiong2022,Manne1960,Jain1999}. Its variants arise regularly in industry \cite{Liu2015}, most notably supporting advanced manufacturing \cite{Zhang2019,Meeran2012,Smaili2025}. In its baseline form, there are a set of $J$ jobs that should be manufactured across a set of $M$ machines in a sequence of operations. A common objective is to minimize makespan (i.e., the total length of time from start of production to completion) \cite{Jain1999}. Alternative objectives (e.g., minimize mean flow time \cite{Baykasoglu2017}, energy \cite{Masmoudi2019}, and workload \cite{Kato2018}) allow companies to tailor to their particular setting. Its many variants include the Just-in-Time JSSP (JIT-JSSP) in which each task has a due date \cite{Ahmadian2021-me}; the Flexible JSSP (FJSSP) \cite{Dauzere-Peres2024,Chaudhry2016} in which a task can be accomplished on multiple machines; and the dynamic JSSP in which jobs arrive after production begins \cite{Kundakci2016}. 

JSSP is an NP-hard problem \cite{Garey1976}, and additional jobs increase the complexity of the problem exponentially \cite{Brucker2007}. Exact solution approaches have focused on branch-and-bound or disjunctive methods \cite{Pinedo2022,Brucker1994,Ku2016}. In practice, heuristics are commonly used \cite{Waschneck2016} because of the problem complexity and computational time. These include the shifting bottleneck \cite{Pinedo2022}, local search \cite{Vaessens1996}, or simple dispatching rules (e.g., shortest processing time \cite{Smith1956,Smaili2025}). Metaheuristic methods are also widespread (e.g., genetic algorithms \cite{Candido1998,Tutumlu2023} and Tabu search \cite{DellAmico1993,Xie2023}).

While classical heuristics can yield high-quality solutions for the JSSP, its NP-hard nature motivates the exploration of alternative computational paradigms, such as quantum computing. Quantum approaches for solving the JSSP can be broadly categorized by their hardware paradigm. Quantum annealing has been used to solve QUBO formulations of both the standard JSSP \cite{Pakhomchik2022-aw, Carugno2022-zx, Venturelli2015-ex, Schmidbauer2025-if} and the more complex FJSSP \cite{Denkena2021-vk, Schworm2022-cw, Schworm2024-pd}. In gate-based quantum computing, most prior research relied on variational quantum algorithms (VQAs)  \cite{Sun2024-av, Kurowski2023-bl, Amaro2022-or}; however, non-variational approaches have also been explored, e.g., digitized adiabatic evolution in Ref.~\cite{Dalal2024-yl}. Within this paradigm, a significant focus has also been placed on developing more efficient encodings to reduce the resource requirements for near-term hardware \cite{Leidreiter2024-en, Schmid2024-ir, Bourreau2024-zd}.

Among the various gate-based methods, the Quantum Approximate Optimization Algorithm (QAOA) has emerged as a leading hybrid quantum-classical method for tackling combinatorial optimization problems \cite{Farhi2014-fn}. Conceptually, QAOA can be viewed as a time-discretized form of adiabatic quantum computing. Its relatively simple circuit structure makes it well-suited for implementation on Noisy Intermediate-Scale Quantum (NISQ) devices, and there is evidence that, for certain problem classes, it may offer an advantage over classical algorithms \cite{Farhi2016-ch, Montanaro2024-zh, Shaydulin2024-mf}. Nevertheless, QAOA lacks general performance guarantees, and the classical optimization of its parameters can be a significant challenge \cite{McClean2018-lw}.

A key determinant of performance in VQAs, such as QAOA, is the choice of initial state. A well-chosen starting point can reduce convergence time, help avoid poor local minima during the classical optimization stage, and improve solution quality. A promising strategy for enhancing QAOA  is ``warm-starting'', in which a biased initial state is prepared using prior information rather than starting from a uniform superposition \cite{Egger2021-kw}. Moreover, beyond accelerating convergence, an appropriate initialization may also mitigate training issues such as barren plateaus \cite{Grant2019-am}.

Multiple warm-starting strategies have been proposed. One approach leverages other quantum processes, such as using the output of a quantum annealing run to prepare the initial QAOA state \cite{Willsch2022-cg,Sack2021-kf}. Another employs iterative feedback, introducing adaptive longitudinal bias fields in the mixer Hamiltonian that are dynamically updated during the optimization process \cite{Yu2022-rl,Yu2023-ix, Zhu2022-rl}. A prominent approach draws on classical heuristics, where an approximate classical solution is used to construct a biased initial state that steers the quantum search toward promising regions of the Hilbert space \cite{Egger2021-kw,Tate2023-rt}. This idea has been extended by combining classical warm-starts with engineered mixer Hamiltonians \cite{Carmo2025-dc, Yu2025-bq}. Orthogonal to state initialization, other heuristics seek effective QAOA parameters bypassing the classical optimization stage, for instance through optimized or extrapolated parameter schedules from small problem instances \cite{Zhou2020-gq, Brandao2018-fz, Hess2024-nn, Sakai2025-ag, Shaydulin2021-pi, Kremenetski2023-vy, Montanez-Barrera2025-im, Dehn2025-jm, Shaydulin2023-yl, Wurtz2021-zi}.

In this work, we introduce a QAOA variant that combines a non-variational heuristic with an iterative warm-starting process. The algorithm leverages a fixed parameter schedule, bypassing the need for a classical optimization loop and any associated barren plateaus. The core of our method is a feedback mechanism that refines the initial state at each iteration by computing a bias-field using only information from previous quantum measurement outcomes \cite{Yuan2025-mm}. A key distinction from prior work is our use of a normalized Boltzmann distribution to assign weights to measurement outcomes for the bias-field update. This technique inherently gives greater significance to results associated with lower energy states.

The remainder of this paper is structured as follows. In \cref{sec:jssp}, we introduce the general mathematical formulation of the JSSP. \cref{sec:quantum_methods} provides an overview of the general approach for mapping combinatorial optimization problems to quantum Hamiltonians, details the construction of our proposed algorithm, and introduces the variational algorithm used for comparison. In \cref{sec:results}, we describe the specific JSSP problem formulation and present the results of applying the algorithms to various problem sizes in both ideal simulations and on quantum hardware. Finally, in \cref{sec:outlook}, we summarize our findings and discuss avenues for future research.

\section{Job Shop Scheduling Problem}
\label{sec:jssp}
\noindent
The Job Shop Scheduling Problem (JSSP) \cite{Manne1960} is a classical combinatorial optimization problem renowned for its computational complexity and broad industrial relevance. The problem involves scheduling a set of $J$ jobs, denoted as $\mathcal{J} = \{j_1, \ldots, j_J\}$, on a set of $M$ machines, $\mathcal{M} = \{m_1, \ldots, m_M\}$. Each job $j$ consists of a sequence of operations that must be executed in a specific order. Each operation has a pre-assigned machine and a fixed processing time. The primary objective is to find a schedule that minimizes the makespan, i.e., the total time required to complete all jobs, subject to two main constraints: each machine can only process one operation at a time, and the precedence order of operations for each job must be respected.

A common approach to solve the JSSP on a quantum computer is to first map it to a Quadratic Unconstrained Binary Optimization (QUBO) problem. Following the time-indexed representation approach \cite{Venturelli2015-ex, Kurowski2023-bl}, the schedule is defined using a set of binary variables $x_{mjt}$, where the indices represent the machine, job, and start time slot, respectively
\begin{align}
    x_{mjt} =
    \begin{cases}
        1 & \text{if the operation of job } j \text{ on} \\ 
        & \text{machine } m \text{ starts at time } t, \\
        0 & \text{otherwise.}
    \end{cases}
    \label{eq:x_mjt}
\end{align}

This variable definition presumes that for any given job $j$, there is at most one operation that runs on a specific machine $m$, which holds for the standard JSSP.

The time index $t$ is discretized and constrained by a deadline $T$, which represents the maximum allowable time for all jobs to complete. $T$ is bounded and can be estimated heuristically. Let $p_{mj}$ be the processing time of the operation for job $j$ on machine $m$. The upper bound on the makespan, $T_{\text{max}}$, is the total processing time of all operations
\begin{equation}
    T_{\text{max}} = \sum_{j \in \mathcal{J}} \sum_{m \in \mathcal{M}_j} p_{mj},
\end{equation}
where $\mathcal{M}_j$ is the set of machines that job $j$ must visit. The lower bound $T_{\text{min}}$ is the time required for the single longest job
\begin{equation}
    T_{\text{min}} = \max_{j \in \mathcal{J}} \sum_{m \in \mathcal{M}_j} p_{mj}.
\end{equation}

The JSSP constraints are enforced by constructing a cost function $C(\mathbf{x})$ composed of penalty terms. A valid schedule must satisfy three primary constraints: (1) each operation must be scheduled exactly once (job assignment constraint); (2) no two operations can be processed on the same machine simultaneously (time assignment constraint); and (3) the original precedence order of operations within each job must be preserved (process order constraint). A schedule is feasible if and only if all of these penalty terms are zero. To guide the search towards solutions with small makespans, an objective term is added that penalizes late finishing times. The final cost function is a sum of these penalty and objective terms
\begin{equation}
    C(\mathbf{x}) = \lambda_1 c_1(\mathbf{x}) + \lambda_2 c_2(\mathbf{x}) + \lambda_3 c_3(\mathbf{x}) + c_{\text{obj}}(\mathbf{x}),
    \label{eq:jit-jssp_cost_function}
\end{equation}
where $\mathbf{x}$ is the complete set of binary variables and $\{\lambda_i\}$ are penalty weights chosen to be large enough to enforce the constraints. Each term $c_i(\mathbf{x})$ and $c_{\text{obj}}(\mathbf{x})$ is constructed to be at most quadratic in the variables $x_{mjt}$, thus ensuring the entire cost function is a valid QUBO \cite{Kurowski2023-bl}.

\section{Quantum Approach to Combinatorial Optimization}
\label{sec:quantum_methods}
\noindent
Having formulated the JSSP as a QUBO problem, the goal is to find the binary assignment $\mathbf{x}^\star$ that minimizes the cost function in \cref{eq:jit-jssp_cost_function}. In the quantum computing paradigm, this is achieved by finding the ground state of a corresponding Ising Hamiltonian. To map the classical QUBO cost function $C(\mathbf{x})$ to a quantum Hamiltonian $H_C$, each binary variable $x_k$ in the vector $\mathbf{x}$ is promoted to a quantum operator acting on a single qubit via the transformation $x_k \to (1 - \sigma^k_z)/2$, where $\sigma^k_z$ is the Pauli-$Z$ matrix acting on the $k$-th qubit. This converts the cost function into a diagonal Hamiltonian of the form
\begin{equation}
    H_C = C\left(x_k \to \frac{1 - \sigma^k_z}{2}\right).
    \label{eq:hamiltonian}
\end{equation}
In this representation, the optimal solution to the JSSP corresponds to the ground state of $H_C$.

The predominant strategy for finding this ground state on current generation of NISQ hardware has been through VQAs. In a VQA, one prepares a parameterized trial state (ansatz) $|\psi(\boldsymbol{\theta})\rangle$, on the quantum computer. A classical optimizer then iteratively updates the real-valued parameters $\boldsymbol{\theta}$ to minimize the expectation value of the Hamiltonian, $\mel{\psi(\boldsymbol{\theta})}{H_C}{\psi(\boldsymbol{\theta})}$. This iterative quantum-classical optimization loop characteristic of VQAs can suffer from challenges such as ``barren plateaus'', which are regions in the parameter landscape where cost function gradients with respect to the variational parameters vanish exponentially with increasing problem size \cite{McClean2018-lw}, making the classical optimization intractable. Furthermore, performance is highly dependent on the choice of the ansatz, and the optimization can become trapped in poor local minima.

To address these challenges, in this work we introduce the Iterative-QAOA, a non-variational variant of the standard QAOA, and benchmark its performance against the Variational Quantum Imaginary Time Evolution (VarQITE) for solving the JSSP. 

\subsection{Iterative-QAOA}
\label{sec:iterqaoa}

\begin{figure*}[t!]
    \centering
    \includegraphics[width=\textwidth]{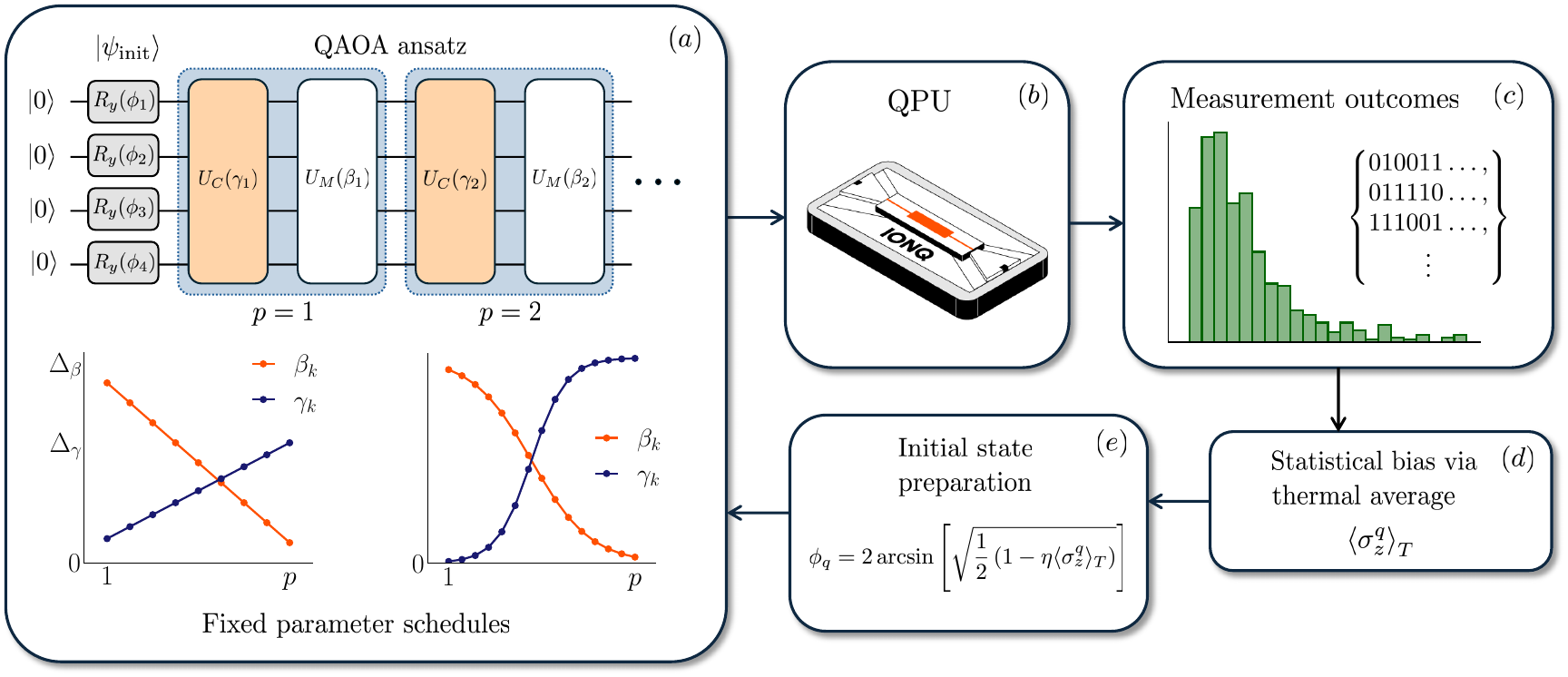}
    \caption{Workflow of the Iterative-QAOA algorithm. $(a)$ A non-variational QAOA ansatz is prepared using a fixed number of layers $p$ and a predetermined schedule for the angles $\{\beta_k\}$ and $\{\gamma_k\}$. The circuit applies alternating cost ($U_C(\gamma_k)=\exp[-i\gamma_kH_C]$) and mixer ($U_M(\beta_k)=\exp[-i\beta_kH_M]$) unitaries to an initial state $|\psi_\text{init}^{(j)}\rangle$. $(b)$ This circuit is executed on quantum hardware (QPU) yielding $(c)$ a set of classical bitstrings. $(d)$ These measurement outcomes are used to compute a statistical bias for each qubit, given by the thermal expectation value $\langle \sigma^q_z \rangle_{T}$. $(e)$ The computed bias determines the angles $\{\phi_q\}$ of the next iteration's initial state $|\psi_\text{init}^{(j+1)}\rangle$, which is prepared by a layer of single-qubit rotations $R_y(\phi_q)$.}
    \label{fig:iterqaoa_diagram}
\end{figure*}

\noindent
The Iterative-QAOA algorithm is built upon the standard QAOA \cite{Farhi2014-fn, Farhi2016-ch} framework, which we briefly summarize here. The algorithm begins by encoding the problem into a cost Hamiltonian $H_C$ in \cref{eq:hamiltonian}, whose ground state corresponds to the optimal solution. A parameterized quantum state, or ansatz, is then prepared on a quantum computer. Starting from an initial state $|\psi_\text{init}\rangle$, typically the uniform superposition $|+\rangle^{\otimes N}$, the $p$-layer QAOA ansatz is constructed by alternately applying unitary operators derived from the cost Hamiltonian $H_C$ and a mixer Hamiltonian $H_M$. A common choice for the mixer, is the transverse field Hamiltonian, $H_M = \sum_{i=1}^{N} \sigma_x^i$. The full ansatz is then given by
\begin{equation}
    |\psi_p(\boldsymbol{\gamma}, \boldsymbol{\beta})\rangle = \prod_{k=1}^{p} e^{-i\beta_k H_M} e^{-i\gamma_k H_C} |\psi_\text{init}\rangle,
    \label{eq:qaoa_ansatz}
\end{equation}
where $\boldsymbol{\gamma} = (\gamma_1, \ldots, \gamma_p)$ and $\boldsymbol{\beta} = (\beta_1, \ldots, \beta_p)$ are $2p$ real-valued variational parameters. A classical optimizer is then used to find the optimal parameters $(\boldsymbol{\gamma}^\star, \boldsymbol{\beta}^\star)$ by minimizing the expectation value of the cost Hamiltonian $H_C$. However, as mentioned earlier, this quantum-classical process can be resource-intensive and is susceptible to challenges common to VQAs.

To circumvent this often costly classical optimization, several non-variational QAOA strategies have been proposed. In these approaches, the parameters $(\boldsymbol{\gamma}, \boldsymbol{\beta})$ are determined by a pre-defined schedule rather than being variationally optimized. Such schedules are frequently inspired by insights from quantum annealing and adiabatic evolution \cite{Brandao2018-fz, Hess2024-nn, Sakai2025-ag}, where the QAOA procedure is viewed as a digitized version of an analog process \cite{Willsch2022-cg}; or by exploiting symmetries of the classical objective function \cite{Shaydulin2021-pi}. A prime example and arguably the most straightforward schedule is used in the Linear Ramp QAOA (LR-QAOA) \cite{Kremenetski2023-vy, Montanez-Barrera2025-im, Dehn2025-jm}. The LR-QAOA protocol replaces the variational search for optimal angles with a fixed, linear schedule. For a $p$-layer ansatz, this schedule is defined by only two global hyperparameters, $\Delta_\gamma$ and $\Delta_\beta$, which set the range for the respective parameter ramps. The parameters for layer $k$ (for $k = 0, \ldots, p-1$) are then given by
\begin{align}
    \beta_k = \left(1 - \frac{k}{p}\right)\Delta_\beta, \qquad \gamma_k = \frac{k+1}{p}\Delta_\gamma.
    \label{eq:lr_qaoa_schedule}
\end{align}

To enhance the performance of fixed-schedule protocols such as LR-QAOA, we propose an iterative method that refines the algorithm's initial state. Our approach (Iterative-QAOA) periodically updates the starting state based only on measurement outcomes from the previous run \cite{Yuan2025-mm}, while keeping the number of QAOA layers and the parameter schedule fixed. After each run $j$, the set of measurement bitstrings $\{s_j\}$ and their corresponding energies $\{E_j\}$, where $E_j = \mel{s_j}{H_C}{s_j}$, is used to guide the search for the next iteration.

The update mechanism calculates a bias \cite{Cadavid2025-pm} for each qubit $q$ by computing the thermal expectation value of its Pauli-$Z$ operator $\sigma^q_z$, from the measurement outcomes. This is achieved using a Boltzmann distribution over the measured states, $P(s_j) = \exp(-\beta_{T} E_j) / Z$, where $Z = \sum_k \exp(-\beta_{T} E_k)$ is the canonical partition  that normalizes the distribution and $\beta_{T}$ is a hyperparameter, analogous to the inverse temperature of the ensemble. The role of $\beta_T$ here is to control the intensity of the feedback; a large value creates a sharp probability distribution that focuses the update on the lowest-energy states, while a smaller value incorporates a wider sample of outcomes for a more gradual update. The bias for each qubit is then the thermal expectation value of $\sigma^q_z$ with respect to this distribution
\begin{equation}
    \langle \sigma^q_z \rangle_{T} = \sum_i P(s_i) \mel{s_i}{\sigma^q_z}{s_i}.
\end{equation}

From these biases, a new initial product state $|\psi_{\text{init}}^{(j+1)}\rangle$ is prepared for the next iteration ($j+1$). The state for each qubit is updated based on the probability $p_q$ of it being initialized to the $|1\rangle$ state, calculated as
\begin{align}
    p_q = \frac{1}{2}\left(1 - \eta \langle \sigma^q_z \rangle_{T}\right),
\end{align}
where $\eta\in\{-1,1\}$ is a hyperparameter that directs the feedback, either reinforcing $(\eta = +1)$ or reversing $(\eta = -1)$ the trend observed in the low-energy states. The new initial state is a product state of the form
\begin{align}
    |\psi_{\text{init}}^{(j+1)}\rangle = \bigotimes_{q=1}^{N} \left(\sqrt{1-p_q}|0\rangle_q + \sqrt{p_q}|1\rangle_q\right).
\end{align}

This state is efficiently prepared by applying a layer of single-qubit rotations $R_y(\phi_q)$ to the $|0\rangle^{\otimes N}$ state, with rotation angles given by $ \phi_q = 2\arcsin(\sqrt{p_q})$ and the mixer Hamiltonian $H_M$ is suitably modified such that this new initial state is an eigenstate of the mixer \cite{Egger2021-kw}. The algorithm proceeds by repeating this process which progressively directs the search towards areas of the Hilbert space corresponding to optimal solutions. In \cref{fig:iterqaoa_diagram} we summarize the workflow of this algorithm. 

\subsection{Variational Quantum Imaginary Time Evolution}\label{sec:varqite}
\noindent
The Variational Imaginary Time Evolution (VarQITE) algorithm provides an efficient method to approximate the ground state $\ket{\psi_{\mathrm{GS}}}$ of a Hamiltonian $H_C$, such as those encoding combinatorial optimization problems. In imaginary time $\tau \equiv it$, the normalized quantum state evolves according to  
\begin{align}
    \pdv{\tau}\ket{\psi(\tau)} = -(H_C - E_\tau) \ket{\psi(\tau)},
\end{align}
where $E_\tau$ is a normalization factor. For a static Hamiltonian, the formal solution  
\begin{align}
    \ket{\psi(\tau)} = \frac{e^{-\tau H_C}}{\sqrt{\ev{e^{-2\tau H_C}}{\psi(0)}}}\ket{\psi(0)}
\end{align}
projects any initial state, with non-zero ground-state overlap, toward $\ket{\psi_{\mathrm{GS}}}$ as $\tau \to \infty$.

Implementing the non-unitary operator $\exp(-\tau H_C)$ directly on quantum hardware is, however, impractical. Instead, VarQITE approximates this evolution using a parameterized quantum circuit
$\ket{\psi(\btheta)} = U(\btheta)\ket{+}^{\otimes N}$, where $U(\btheta)$ defines a unitary ansatz characterized by real parameters $\btheta = (\theta_1, \theta_2, \ldots, \theta_{N_p})$. The goal is to update these parameters as functions of imaginary time, $\btheta(\tau)$, so that the variational state $\ket{\psi(\btheta(\tau))}$ approximates the exact imaginary-time-evolved state. Rather than using the McLachlan variational principle \cite{Yuan2019-oj,McArdle2019-zp}, which requires $\mathcal{O}(N_p^2)$ circuit evaluations per step, we employ the more efficient formulation of Ref.~\cite{Morris2024-gj}, which enforces the dynamics of each Pauli term $P_i$ in $H_C$ through 
\begin{align}
\nonumber
    \pdv{\expval{P_i}}{\tau} &= \sum_j 2\Re\left[ \bra{\psi(\btheta)} P_i\, \pdv{\theta_j}\ket{\psi(\btheta)}\right] \dot{\theta}_j\\
    &= -\mel{ \psi(\btheta)}{\{P_i,H_C-E_{\tau}\}}{\psi(\btheta)},
\end{align}
yielding the linear system of the form $\sum_jA_{ij} \dot{\theta}_j = B_i$, where
\begin{align}
    \nonumber
    A_{ij} = \Re\left[ \bra{\psi(\btheta)} P_{i}\, \pdv{\theta_j}\ket{\psi(\btheta)}\right], \\
    \qquad B_i = -\frac{1}{2}\mel{ \psi(\btheta)}{\{P_i,H_C-E_{\tau}\}}{\psi(\btheta)}.
\end{align}

Each column of $A$ is obtained from two parameter-shift measurements \cite{Mitarai2018-mu,Schuld2019-cq}, while for QUBO Hamiltonians containing only commuting Pauli-$Z$ terms, all elements of $B$ can be evaluated from a single circuit. The parameters are then evolved in imaginary time via a first-order update rule,  $\btheta \to \btheta + \Delta\tau \dot{\btheta}$, where $\dot{\btheta}$ is computed by solving the above linear system.

This formulation requires only $2N_p+1$ circuit evaluations per time step, significantly improving efficiency over previous implementations while retaining the key physics of imaginary-time evolution. It thus offers a practical and potentially scalable approach for ground-state preparation on near-term quantum hardware.

The imaginary time step $\Delta\tau$ is a critical hyperparameter governing the algorithm's convergence and accuracy. Its selection involves a trade-off between fidelity and efficiency; a small step size improves the accuracy of the Euler integration at the cost of more iterations, whereas a large step size can lead to numerical instability and hinder convergence. While a constant step size is the simplest approach, we find that using a step-size schedule $\Delta\tau(\tau)$ yields substantially better performance. This suggests that adaptive strategies, where $\Delta\tau$ is adjusted dynamically based on the system's state, could further enhance the algorithm's efficiency and robustness.

\section{Results and Discussion}
\label{sec:results}

\noindent
In this section, we present a comprehensive evaluation of the Iterative-QAOA algorithm applied to instances of a specific JSSP variant. We evaluate its performance against two key reference algorithms: the underlying non-variational LR-QAOA establishing the fixed parameter schedule, and VarQITE as a representative of the variational algorithms class. We begin by detailing the specific JSSP cost function used. The algorithms are then evaluated across a range of problem sizes (see \cref{sec:jssp-inst-gen} for details on problem instances) using both ideal noiseless simulations and executions on a trapped-ion quantum processor (see \cref{sec:ion-traps-qpus} for technical details of the quantum hardware) to assess the algorithms' performance and robustness to hardware noise. We conclude with an investigation of the method's scaling potential using large-scale simulations up to $\sim100$ qubits.

\subsection{JSSP Formulation}
\label{sec:probl_formulation}

\noindent
Following the formulation proposed in Ref.~\cite{Amaro2022-or}, we consider a JIT-JSSP variant tailored to a steel manufacturing process in which due dates are defined for each job, and the objective is to minimize deviations from the due dates and changeover (switching) costs.

In this problem, a set of $J$ jobs must each be processed on $M$ machines in a fixed order with fixed idle slots. The schedule of each machine is divided into $T$ consecutive time slots, and each non-idle time slot must be assigned to exactly one job. The processing times for all job operations are assumed to be equal. To represent the schedule, we introduce binary decision variables $x_{mjt}\in\{0,1\}$ that indicate whether job $j$ is processed by machine $m$ at time $t$. 

Each job $j$ has a due time $d_j$, and finishing before or after this time incurs a penalty that scales linearly with the deviation. The early/late delivery cost for job $j$ can be formulated as follows
\begin{align}
    \nonumber
    u_j(\mathbf{x}) & = c_e \sum_{t=1}^{d_j} (d_j - t)x_{Mjt} \\
    \nonumber
    & + c_l \sum_{t=d_j+1}^{T} (t - d_j)x_{Mjt}\\
    & \forall j = 1,...,J,
\end{align}
where $c_e$ and $c_l$ are constants controlling the magnitude of the early and late delivery penalties, respectively.

Each operation is associated with a production group, and switching production groups in consecutive time slots on the same machine incurs a switching cost. Let $P_{mj}$ denote the production group of job $j$ on machine $m$, and define $G_{j_1j_2}^{(m)} = 1$ if $P_{mj_1}\neq P_{mj_2}$ and 0 otherwise. The production switching cost for machine $m$ is then given by
\begin{align}
\nonumber
    s_m(\mathbf{x}) = c_p \sum_{j_1,j_2=1}^J \sum_{t=1}^{T_m-1} G_{j_1j_2}^{(m)} x_{mj_1t} x_{mj_2(t+1)} \\
    \forall m=1,\dots,M,
\end{align}
where $c_p$ is the production switching cost coefficient. The total cost for a given schedule is then  
\begin{align}
    c(\mathbf{x}) = \sum_{j=1}^J u_j(\mathbf{x}) + \sum_{m=1}^M s_m(\mathbf{x}).
\end{align}
The schedule of each machine comprises a tight sequence of $J$ jobs within the total number of time slots $T_m$, on machine $m$, and feasibility is ensured by the following three sets of constraints. \textit{Job assignment constraints} ensure that each job $j$ is scheduled exactly once on each machine $m$:
\begin{align}
\nonumber
    g_{mj}(\mathbf{x}) = \sum_{t=1}^{T_m} x_{mjt} = 1\\
    \forall m = 1,\dots,M; \ j = 1,\dots,J.
\end{align}

The \textit{time assignment constraints} limit each time slot $t$ on machine $m$ to its appropriate capacity. Let
\begin{align}
\nonumber
    \ell_{mt}(\mathbf{x}) = \sum_{j=1}^J x_{mjt} \\
    \forall m = 1,\dots,M; \ t = 1,\dots,T_m,
\end{align}
and let $A_m$ and $I_m$ denote, respectively, the sets of active and idle slots on machine $m$, with $A_m \cup I_m = \{1, \ldots, T_m\}$ and  $A_m\cap I_m = \emptyset$. 
We impose
\begin{align}
    \ell_{mt}(\mathbf{x}) = \begin{cases}
        1, & \forall m = 1,\dots,M; \ t\in A_m, \\
        0, & \forall m = 1,\dots,M; \ t\in I_m.
    \end{cases}
\end{align}

Finally, the \textit{process order constraints} enforce that jobs must progress through the machines in the correct order, preventing any job from appearing earlier on a later machine or appearing on multiple machines in the same time slot:
\begin{align}
    \nonumber
    q_{mj}(\mathbf{x}) = \sum_{t=2}^{J} \sum_{t'=1}^{t} x_{mjt}x_{(m+1)jt'} = 0, \\
    \forall m=1,\dots,M-1;\; j=1,\dots,J.
    \label{eq:proc_order_constr}
\end{align}
To incorporate these constraints within a quantum framework, we convert the total cost function into a QUBO form by adding penalty terms yielding
\begin{align}
    \nonumber
    C(\mathbf{x}) & = c(\mathbf{x}) + \lambda \sum_{m=1}^M \sum_{j=1}^J \left(g_{mj}(\mathbf{x}) - 1\right)^2 \\ 
    \nonumber
    & + \lambda \sum_{m=1}^M \left[ \sum_{t\in A_m} \left(\ell_{mt}(\mathbf{x}) - 1\right)^2 + \sum_{t\in I_m} \ell_{mt}(\mathbf{x}) ^2 \right]\\
    &+ \lambda \sum_{m=1}^{M-1} \sum_{j=1}^J q_{mj}(\mathbf{x}),
    \label{eq:jit-jssp_cost_func}
\end{align}
where $\lambda>0$ is a penalty weight. In this formulation, the idle slots are treated as fixed zero-capacity positions.

This formulation closely follows that of Ref.~\cite{Amaro2022-or}, with only two key modifications. First, we enforce a stricter sequential processing constraint, requiring a job to finish on one machine before starting on the next. This is reflected in the upper limit of the inner sum in the process order constraint (\cref{eq:proc_order_constr}), which changes from $t-1$ to $t$. Second, we treat idle slots ($I_m$) as fixed parameters in the problem definition, rather than handling them as optimizable dummy variables as is done in Ref.~\cite{Amaro2022-or}. This modification simplifies the model by reducing the total number of variables.

It is worth noting that this setup defines due dates at the job level. More complex variants of the JIT-JSSP exist where due dates apply at the operation level, leading to significantly higher complexity and greater difficulty in obtaining optimal solutions \cite{Ahmadian2021-me}.

\begin{figure*}[t!]
    \centering
    \includegraphics[width=\textwidth]{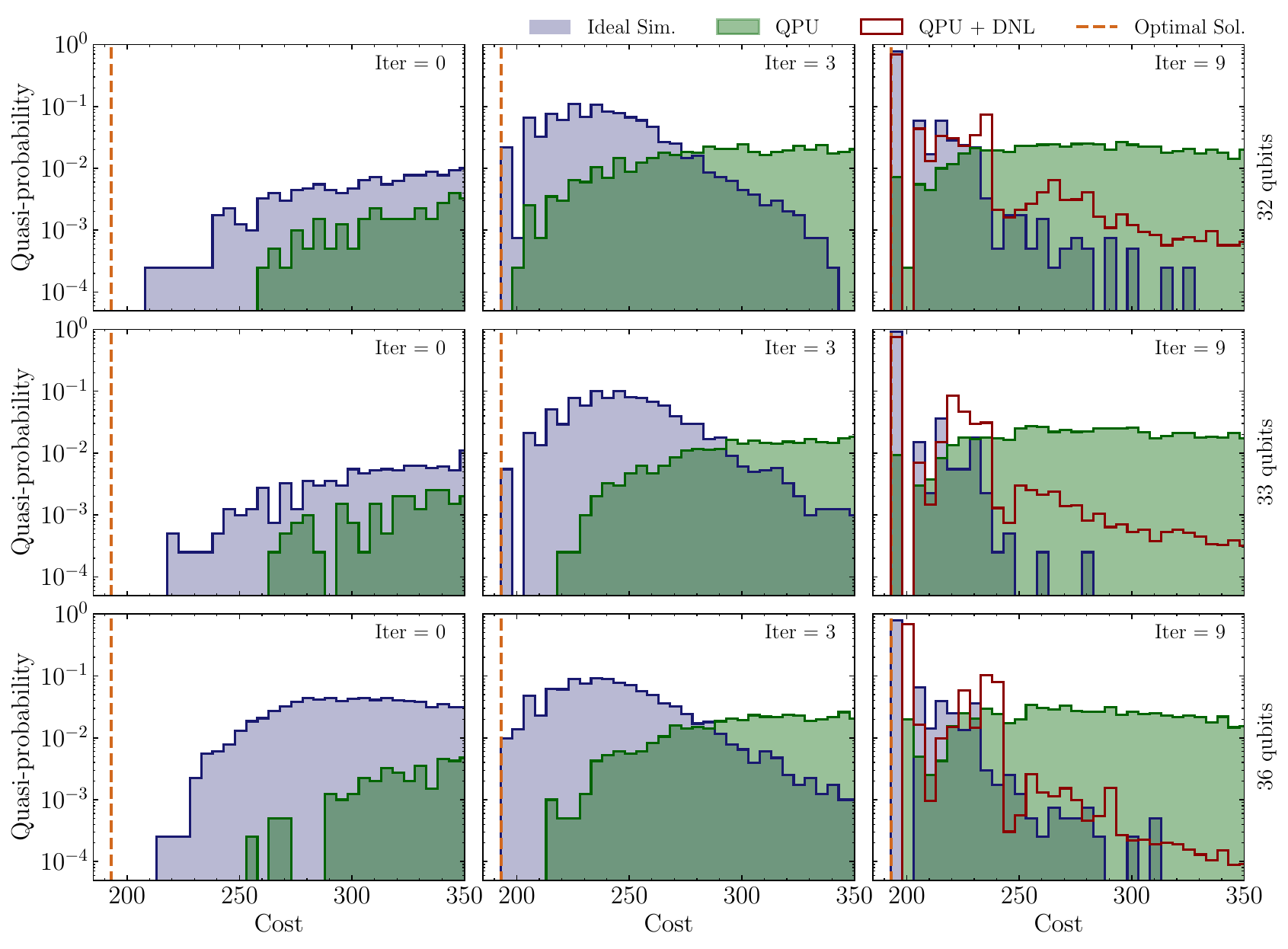}
    \caption{Performance of Iterative-QAOA on JSSP instances of two different sizes executed on IonQ Forte Generation QPU. Shown here are the low-energy sectors of the cost probability distributions for the 32- (top row), 33- (middle row), and 36- (bottom row) qubit instances. The columns display the distribution's evolution at the initial (Iter = 0), an intermediate (Iter = 3), and the final (Iter = 9) iteration. Each panel compares the ideal noiseless simulation results with raw data from the QPU executions and the error-mitigated results after applying DNL. The dashed vertical line indicates the known optimal cost. The algorithm parameters used are $\Delta_{\beta/\gamma} = 0.17$ and a quadratic schedule for $\beta_T\in[0.1,1]$. }
    \label{fig:iterqaoa_qpu-vs-ideal}
\end{figure*}

The JIT-JSSP instances evaluated in this work originate from a master problem comprising $J=20$ jobs and $M=3$ machines. Using the time-indexed formulation with binary variables $x_{mjt}$ (\cref{eq:x_mjt}), the total number of variables for the full instance is $n_\text{var}=\sum_{m=1}^{M} J T_m = 1,300$, where $T_m$ denotes the number of time slots per machine $m$, which in this case is $T_m=\{20, 22, 23\}$ for $m\in\{1, 2, 3\}$ respectively. In our quantum approach, a one-hot encoding (\cref{eq:hamiltonian}) is utilized to map each binary variable to a qubit. Solving the full 1,300-variable instance via quantum simulation or on current quantum hardware is intractable. However, state-of-the-art classical solvers are capable of handling instances of this size (see \cref{sec:classical_solvers}). As detailed in \cref{sec:jssp-inst-gen}, the full instance was solved classically, yielding an optimal makespan of 193. The sub-instances were then constructed by fixing (``freezing'') most variables to the optimal solution, allowing us to generate problems ranging from 24 to 36 qubits for QPU execution, and larger 50- and 97-qubit instances suitable for tensor network simulations. Hence,  the optimal schedule for all sub-problems corresponds to the classically computed global optimum, with a ground state energy of 193.
The detailed methodology for generating these sub-instances is described in \cref{sec:jssp-inst-gen}, where we also show the specific problem parameters (\cref{tab:prod_groups,tab:exp_param}). 

\subsection{Performance of Iterative-QAOA}

\begin{figure*}[t!]
    \centering
    \includegraphics[width=\textwidth]{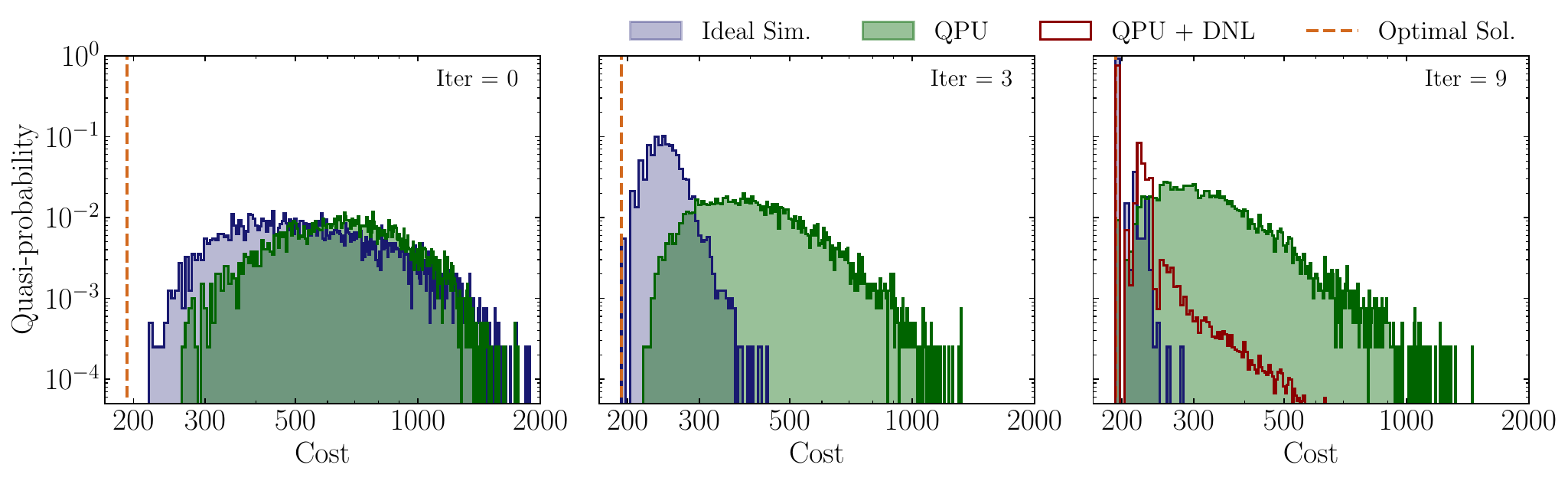}
    \caption{Evolution of the full cost distribution for the 33-qubit JSSP instance under Iterative-QAOA on the IonQ Forte Generation QPU. The panels display the distributions at the initial (Iter = 0), an intermediate (Iter = 3), and the final (Iter = 9) iteration. The algorithm parameters used are $\Delta_{\beta/\gamma} = 0.17$ and a quadratic schedule for $\beta_T\in[0.1,1]$.}
    \label{fig:iterqaoa_qpu-vs-ideal_full}
\end{figure*}

\noindent
We evaluate the performance of our Iterative-QAOA algorithm on JIT-JSSP instances with 32, 33, and 36 qubits (see \cref{fig:gantt_33qubit}). The quantum circuit within each iteration is based on the LR-QAOA schedule (see \cref{eq:lr_qaoa_schedule}), with an ansatz depth of $p=5$ for the 32- and 33-qubit instances and $p=6$ for the 36-qubit instance. Based on an extensive parameter search detailed in \cref{sec:lrqaoa_param_exploration}, an optimal ramp parameter of $\Delta_{\beta/\gamma}=0.17$ was used for all instances. For the feedback loop that creates the updated initial state at each iteration, the inverse temperature hyperparameter $\beta_T$ was set to a quadratic schedule that grows from 0.1 to 1.0 over 10 iterations. All runs were executed with 4,000 measurement shots per circuit both on the ideal quantum simulator as well as on quantum hardware. The two smallest instances were simulated using a statevector simulator while the ideal results for the 36-qubit instance were obtained using a Matrix Product State (MPS) \cite{Vidal2003-sf,Schollwock2011-pf} simulator with a maximum bond dimension of $\chi=256$. 

\begin{figure*}[t!]
    \centering
    \includegraphics[width=\textwidth]{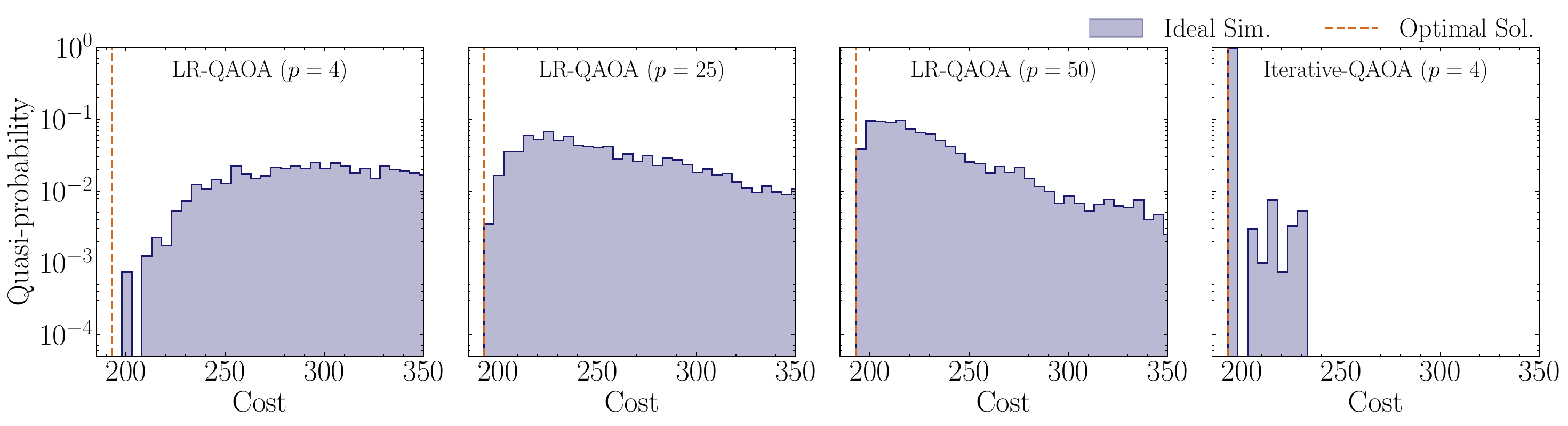}
    \caption{Comparison of LR-QAOA and Iterative-QAOA performance on the 24-qubit JIT-JSSP instance. All panels show the low-energy sector of the final cost distribution from ideal noiseless simulations with 4,000 measurement shots. The first three panels (from left to right) display the results for a standard LR-QAOA circuit with increasing depth of $p=4, \,25,$ and $50$ layers, respectively. The results show that even at a significant depth of $p=50$, the probability of sampling the ground state remains low ($<4\%$). The rightmost panel shows the distribution for $p=4$ Iterative-QAOA after 10 iterations, resulting in a ground state population of 97.92\%. The algorithm parameters used are $\Delta_{\beta/\gamma} = 0.17$ and a quadratic schedule for $\beta_T\in[0.1,1]$.}
    \label{fig:lrqaoa_vs_iterqaoa}
\end{figure*}

The evolution of the algorithm's performance is shown in \cref{fig:iterqaoa_qpu-vs-ideal}, which displays the low-energy sector of the cost distribution at initial, intermediate, and final iterations. In ideal noiseless simulations, the algorithm proved to be highly effective. For all three problem instances, the iterative process converged successfully, producing a final distribution where the probability of sampling the optimal solution (i.e. the ground state) exceeded 78\%. 

\begin{figure*}[t!]
    \centering
    \includegraphics[width=\textwidth]{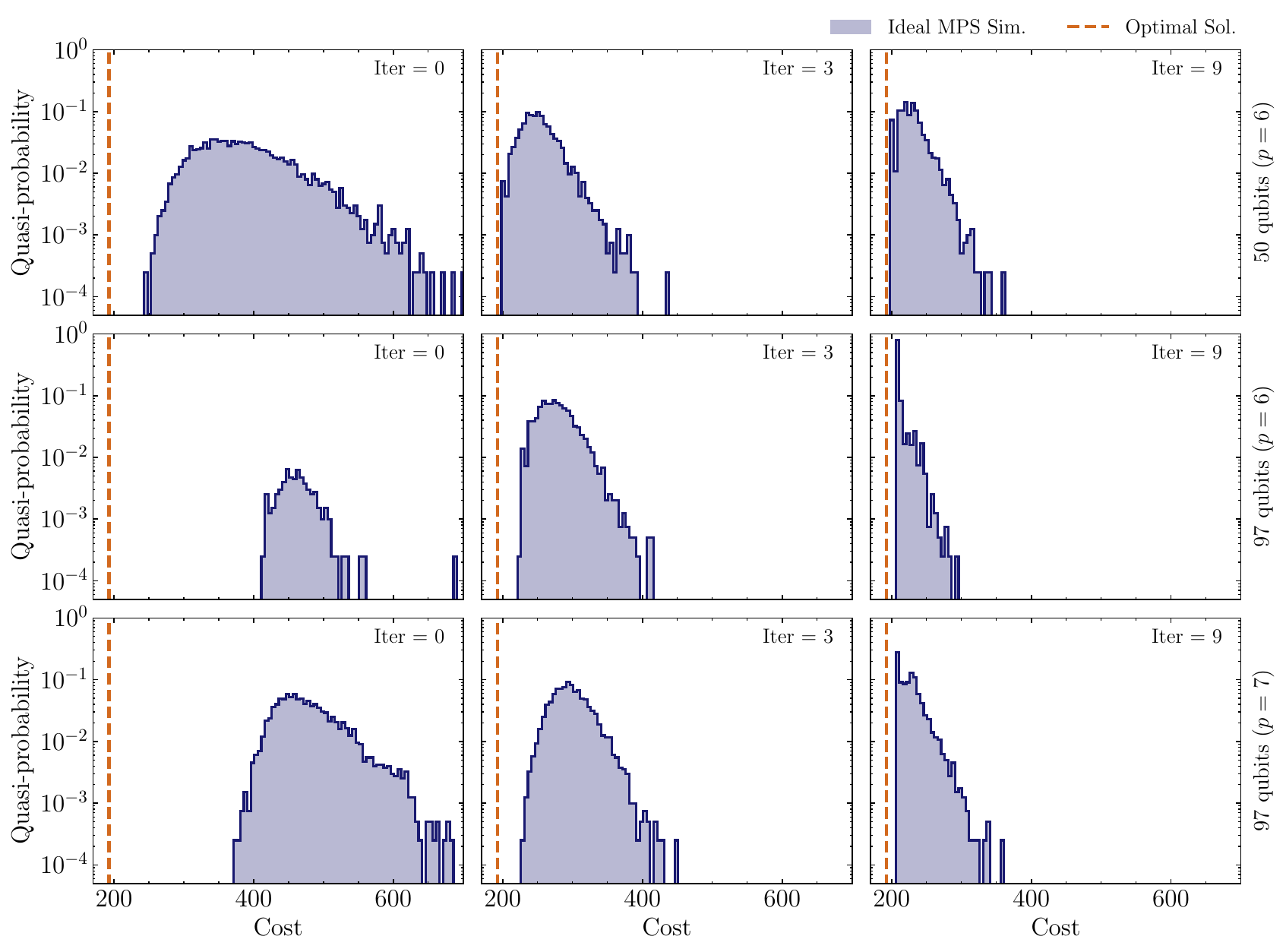}
    \caption{Performance of Iterative-QAOA on 50- and 97-qubit JIT-JSSP instances with an ideal MPS simulator and bond dimension $\chi=256$. Each panel shows the low-energy sectors of the cost probability distributions at the initial (Iter = 0), an intermediate (Iter = 3), and the final (Iter = 9) iteration. The number of layers for the 50-qubit instance was $p=6$ while for the 97-qubit instance we show the two cases of  $p=6$ and 7. The algorithm parameters used are $\Delta_{\beta/\gamma} = 0.17$ and a quadratic schedule for $\beta_T\in[0.1,1]$.}
    \label{fig:iIterative-QAOA_large-inst}
\end{figure*}

On the QPU, the effects of hardware noise become more apparent as the algorithm evolves. This is further illustrated in \cref{fig:iterqaoa_qpu-vs-ideal_full}, which shows the full cost distributions of the 33-qubit instance. In the ideal case, the algorithm rapidly ``squeezes'' the probability distribution towards the ground state, resulting in a final distribution highly skewed towards the optimal solution. The evolution on the hardware is qualitatively similar but less pronounced, indicating that noise effectively broadens the final distribution reducing the probability of sampling the ground state.

To mitigate the impact of hardware induced errors on the QPU executions, we applied a debiasing with non-linear filtering (DNL) post-processing technique to the raw measurement data (see \cref{sec:dnl} for more details). In both \cref{fig:iterqaoa_qpu-vs-ideal,fig:iterqaoa_qpu-vs-ideal_full}, the DNL technique effectively suppresses the high-energy tails of the noisy distribution while enhancing the probability of the target low-energy states. Nevertheless, even in the raw, unmitigated case, the algorithm demonstrated sufficient robustness to find the optimal solution for the 32- and 33-qubit instances and a high-quality sub-optimal solution, corresponding to the first excited state, for the 36-qubit problem.

To further probe the robustness of the algorithm, we repeated the experiments using a deliberately sub-optimal parameter configuration. We selected a ramp parameter of $\Delta_{\beta/\gamma}=1.25$, a value far from the optimal region (see \cref{sec:lrqaoa_param_exploration}), and a fixed  $\beta_T=0.5$. As detailed in \cref{sec:sub-optimal_lrqaoa}, we observed that the iterative warm-starting process was powerful enough to guide the less effective QAOA ansatz toward the optimal solution in both ideal simulations and hardware runs.

Moreover, we compared the performance of Iterative-QAOA against the underlying LR-QAOA. For this comparison, we used a 24-qubit JIT-JSSP instance with an optimized ramp parameter $\Delta_{\beta/\gamma}=0.17$ (see \cref{sec:lrqaoa_param_exploration}). The noiseless simulations were performed using 4,000 measurement shots.

The results are summarized in \cref{fig:lrqaoa_vs_iterqaoa}. For LR-QAOA, a shallow circuit with $p=4$ layers did not sample the optimal solution. Increasing the depth to $p=25$ and $p=50$ resulted in ground state populations of 0.35\% and 3.82\%, respectively. In contrast, after only 10 iterations, the Iterative-QAOA protocol using a fixed depth of just $p=4$ layers achieved a ground state sampling probability of approximately 97.92\%.

This performance difference highlights the advantages of an efficient warm-starting process, particularly for current and near-term quantum hardware where errors accumulate rapidly with circuit depth. The $p=50$ LR-QAOA circuit requires 4,824 single-qubit and 3,900 two-qubit gates. The shallow $p=4$ circuit used in each Iterative-QAOA run, however, requires only 408 single-qubit and 312 two-qubit gates; an order of magnitude fewer resources per iteration. This makes the iterative approach significantly more robust and practical for execution on current devices.

Furthermore, the scaling of LR-QAOA presents a challenge for larger problem instances. As detailed in \cref{sec:lrqaoa_param_exploration}, the circuit depth required for LR-QAOA to achieve a certain solution quality is expected to grow with the problem size, reaching prohibitive gate counts, particularly for larger instances on near term quantum computers. 

Finally, to probe the algorithm's scaling on instances beyond the capacity of the quantum hardware, we simulated 50- and 97-qubit JIT-JSSP instances. Given the large Hilbert space, as with the 36-qubit instance, these simulations were performed using an MPS simulator with a fixed, maximum bond dimension of $\chi=256$. We used an ansatz depth of $p=6$ for the 50-qubit case, and explored both $p=6$ and $p=7$ for the 97-qubit case, with 4,000 measurement shots for all runs. To test the robustness of our hyperparameter selection, we used the same values of $\Delta_{\beta/\gamma} = 0.17$ and a quadratic schedule for $\beta_T$.

The results, shown in \cref{fig:iIterative-QAOA_large-inst}, indicate that both large-scale instances demonstrated convergence after 10 iterations, with their final probability distributions becoming heavily skewed towards the low-energy sector. Although the true ground state was not sampled in either case, the 50-qubit instance sampled the first excited state of the Hamiltonian. In contrast, the 97-qubit runs converged to a solution of slightly lower quality (4th excited state), indicating increased difficulty of the problem at this scale. One potential reason for this decreased solution quality could be that the circuit depths of $p=6$ and $p=7$ may not be sufficiently expressive to resolve the ground state for problems of this scale. Another possible reason could be that increasing circuit depth also generates more entanglement, requiring a larger bond dimension to maintain accuracy in the MPS approximation, and thus greater classical computational resources. This effect is visible in the 97-qubit case, where increasing the depth from $p=6$ to $p=7$, yielded no improvement in the final energy. This suggests that, at a fixed bond dimension, the simulation could not capture all additional correlations generated by the deeper circuit. Nevertheless, the ability of Iterative-QAOA to consistently find high-quality sub-optimal solutions, even within this constrained scenario, remains a strong indicator of its potential for large-scale problems. All the results discussed in this section are summarized in \cref{tab:results}. 

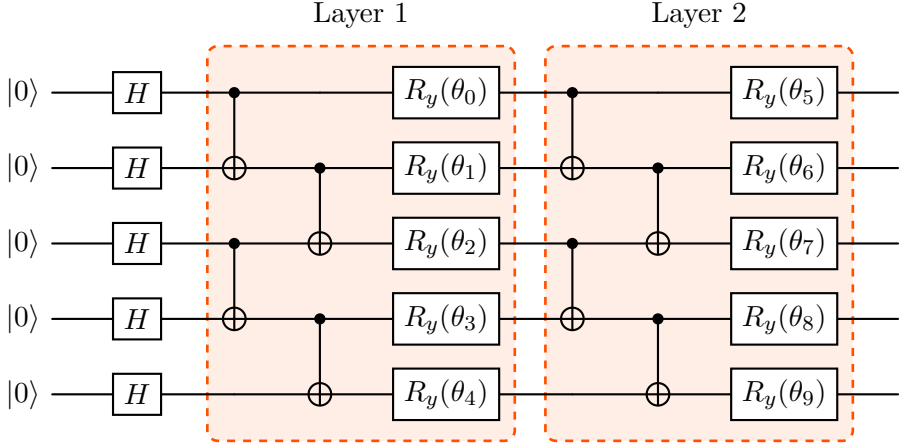
\begin{figure*}[ht!]
    \centering
    \begin{tikzpicture}
        \node[scale=1] {
            \begin{quantikz}[row sep={1cm, between origins}, column sep=0.8cm]
                \lstick{$\ket{0}$} & \gate{H} & \ctrl{1}\gategroup[5,steps=3,style={dashed,rounded
corners, line width=1pt, style=ionq_orange, fill=ionq_orange!10, inner xsep=2pt},background,label style={label
position=above,anchor=north, yshift=0.5cm}]{{Layer 1}} & & \gate{R_y(\theta_0)} & \ctrl{1}\gategroup[5,steps=3,style={dashed,rounded
corners, line width=1pt, style=ionq_orange, fill=ionq_orange!10, inner xsep=2pt},background,label style={label
position=above,anchor=north, yshift=0.5cm}]{{Layer 2}} & & \gate{R_y(\theta_{5})} & \\
                \lstick{$\ket{0}$} & \gate{H} & \targ{} & \ctrl{1} & \gate{R_y(\theta_1)} & \targ{} & \ctrl{1} & \gate{R_y(\theta_{6})} & \\
                \lstick{$\ket{0}$} & \gate{H} & \ctrl{1} & \targ{} & \gate{R_y(\theta_2)} & \ctrl{1} & \targ{} & \gate{R_y(\theta_{7})} & \\      
                \lstick{$\ket{0}$} & \gate{H} & \targ{} & \ctrl{1} & \gate{R_y(\theta_3)} & \targ{} & \ctrl{1} & \gate{R_y(\theta_{8})} & \\
                \lstick{$\ket{0}$} & \gate{H} &  & \targ{} & \gate{R_y(\theta_4)} &  & \targ{} & \gate{R_y(\theta_{9})} &
            \end{quantikz}
        };
    \end{tikzpicture}
    \caption{Structure of the two-layer variational ansatz $\ket{\Psi(\btheta)}$ used for VarQITE runs. The figure shows a simplified 5-qubit version to illustrate the overall layout. The full 32-qubit circuit consists of $N_p = 64$ variational parameters. }
    \label{fig:varqite_ansatz}
\end{figure*}

\begin{figure*}[ht!]
    \centering
    \includegraphics[width=\textwidth]{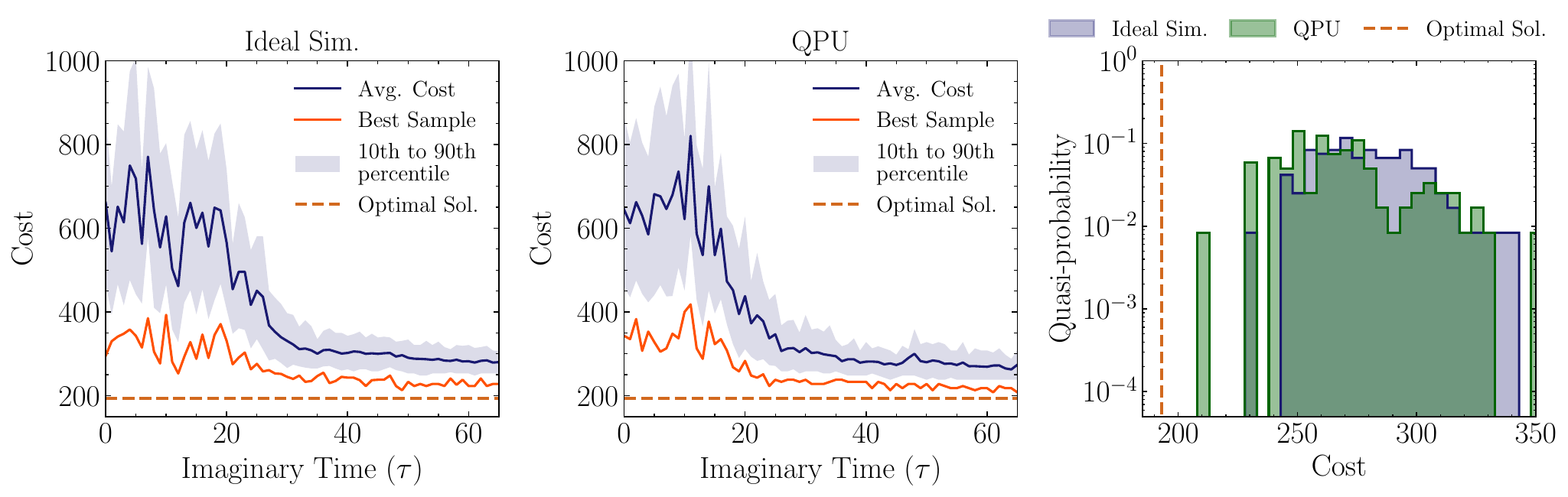}
    \caption{Performance of the VarQITE algorithm on the 32-qubit JIT-JSSP instance. The leftmost plots show the convergence for the ideal noiseless simulation and the QPU execution, respectively, as a function of imaginary time $\tau$. In both convergence plots, the solid blue line corresponds to the average cost $\langle H_C\rangle$, while the shaded region indicates the range between the 10th and 90th percentiles of the sampled distribution. The orange line tracks the cost of the best (lowest-energy) sampled at each time step. The right-most plot displays the low-energy sector of the cost distribution of the last time step.}
    \label{fig:varqite_ideal_vs_qpu}
\end{figure*}

\subsection{Performance of VarQITE}
\label{sec:varqite_results}

\noindent
We also evaluated the performance of the VarQITE algorithm on a representative JIT-JSSP instance of 32 variables, corresponding to a 32-qubit Hamiltonian. For the variational circuit, we employ a two-layer ansatz with nearest-neighbor connectivity and $N_p =64$ variational parameters, as depicted in \cref{fig:varqite_ansatz}. As pointed out in \cref{sec:varqite}, a critical hyperparameter for VarQITE is the imaginary time step $\Delta\tau$. Through preliminary optimization, we identified a variable step-size schedule of the form 
\begin{align}
    \Delta\tau(\tau)=\Delta_0(1-r)^\tau,
\end{align}
to be most effective, using the optimized values $\Delta_0 =0.1$ and $r=0.06$ for the evolution. 

The algorithm was executed for 65 imaginary time steps, at which point convergence was observed. We performed both ideal noiseless simulations and executions on quantum hardware, using 120 measurement shots for each of the 129 circuits required per time step. The results of these runs are portrayed in \cref{fig:varqite_ideal_vs_qpu} and summarized in \cref{tab:results}. The best schedule sampled from the ideal simulation corresponded to a cost of 228, while the QPU execution yielded a lower best cost of 208. In both cases, this best solution was found with a low probability of approximately 0.83\% (corresponding to a single measurement shot). Relative to the optimal value of 193, these costs correspond to the fourteenth and fourth excited states of the problem Hamiltonian, respectively.

Although VarQITE did not reach the ground state for this 32-qubit problem, the results highlight two of its key characteristics. First, the algorithm exhibits a surprising robustness to hardware noise. The overall convergence on the QPU was comparable to the ideal simulation, yet it sampled a single state with a lower cost. Given that this occurred in only a single measurement shot, it is likely due to statistical fluctuation induced by noise. Second, in smaller-scale test runs, the algorithm performed significantly better indicating potential challenges for scaling this algorithm to larger problem instances.

\begin{table*}[!t]
\footnotesize
    \begin{tabularx}{\textwidth}{XLLXLLXLX}
    \toprule
        Algorithm & \makecell[l]{Instance\\Size\\(Qubits)} &\makecell[l]{Circuit\\Layers\\$(p)$} & \makecell[l]{Total Circuit\\Executions}& \makecell[l]{1Q\\Gates} & \makecell[l]{2Q\\Gates}  & Backend & \makecell[l]{Best\\Solution} & \makecell[l]{Opt. Sol.\\Probability}  \\
        \midrule
        \multirow{13}{*}{\makecell{Iterative-\\QAOA}} &  24 &  4 & 40,000 & 408 & 312 & Ideal Sim. & \textbf{193} &  0.979 \\
        \cmidrule{2-9} \\[-1em]
        & \multirow{3}{*}{32} &  \multirow{3}{*}{5} & \multirow{3}{*}{40,000} & \multirow{3}{*}{672} & \multirow{3}{*}{780} & Ideal Sim. & \textbf{193} & 0.780 \\
        & & & & & & QPU &\textbf{193} & 0.007 \\
        & & & & & & QPU+DNL &\textbf{193} & 0.693\\
        \cmidrule{2-9} \\[-1em]
        &  \multirow{3}{*}{33} & \multirow{3}{*}{5} & \multirow{3}{*}{40,000} & \multirow{3}{*}{693} & \multirow{3}{*}{840} & Ideal Sim. & \textbf{193} & 0.914  \\
        & & & & & & QPU & \textbf{193} & 0.009\\
        & & & & & & QPU+DNL &\textbf{193} & 0.755\\
        \cmidrule{2-9} \\[-1em]
        &  \multirow{3}{*}{36} & \multirow{3}{*}{6} & \multirow{3}{*}{40,000} & \multirow{3}{*}{900} & \multirow{3}{*}{972} & \makecell[l]{Ideal Sim.\\ (MPS)} & \textbf{193} &  0.794 \\
        & & & & & & QPU & 198 & 0.000\\
        & & & & & & QPU+DNL & 198 & 0.000\\
        \cmidrule{2-9} \\[-1em]
        &  50 &  6 & 40,000 & 1,250 & 1,686 & \makecell[l]{Ideal Sim.\\ (MPS)} & 198 & 0.000 \\
        \cmidrule{2-9} \\[-1em]
        &  \multirow{2}{*}{97} & 6 & 40,000 & 2,425 & 5,208 & \multirow{2}{*}{\makecell[l]{Ideal Sim.\\ (MPS)}} & 206 & 0.000 \\
        & & 7 & 40,000 & 2,813 & 6,076 & & 206 & 0.000 \\
        \cmidrule{1-9} \\[-1em]
        \multirow{3}{*}{LR-QAOA} & \multirow{3}{*}{24} & 4 & 4,000 & 408 & 312 & \multirow{3}{*}{Ideal Sim.} & 198 & 0.000\\
        & & 25 & 4,000 & 2,424 & 1,950 & & \textbf{193} & 0.004\\
        & & 50 & 4,000 & 4,824 & 3,900 & & \textbf{193} & 0.038\\
        \cmidrule{1-9} \\[-1em]
        \multirow{2}{*}{VarQITE} & \multirow{2}{*}{32} & \multirow{2}{*}{2*} & \multirow{2}{*}{1,006,200}& \multirow{2}{*}{72} & \multirow{2}{*}{46} & Ideal Sim. & 228 & 0.000\\
        & & & & & & QPU & 208 & 0.000\\
        \bottomrule
    \end{tabularx}
    \caption{Summary of performance of Iterative-QAOA, LR-QAOA, and VarQITE across ideal noiseless simulations (statevector or MPS), raw QPU executions, and DNL error-mitigated runs. Resource requirements are quantified by circuit depth $(p)$, gate counts (two-qubit gates are accounted as $R_{ZZ}$ since they are native to this class of hardware), and total circuit executions, which aggregates all shots and iterations to represent the overall workload on the QPU. Performance is measured by the best (lowest-cost) solution obtained for the makespan of the JIT-JSSP problem and the sampling probability of the optimal configuration. The number of layers for VarQITE, marked with an asterisk (*), corresponds to the depth of the variational ansatz defined in \cref{fig:varqite_ansatz}. Optimal solutions are shown in bold.\medskip}
    \label{tab:results}
\end{table*}

\subsection{Discussion}
\label{sec:discussion}

\begin{figure}[t!]
    \centering
    \includegraphics[width=\columnwidth]{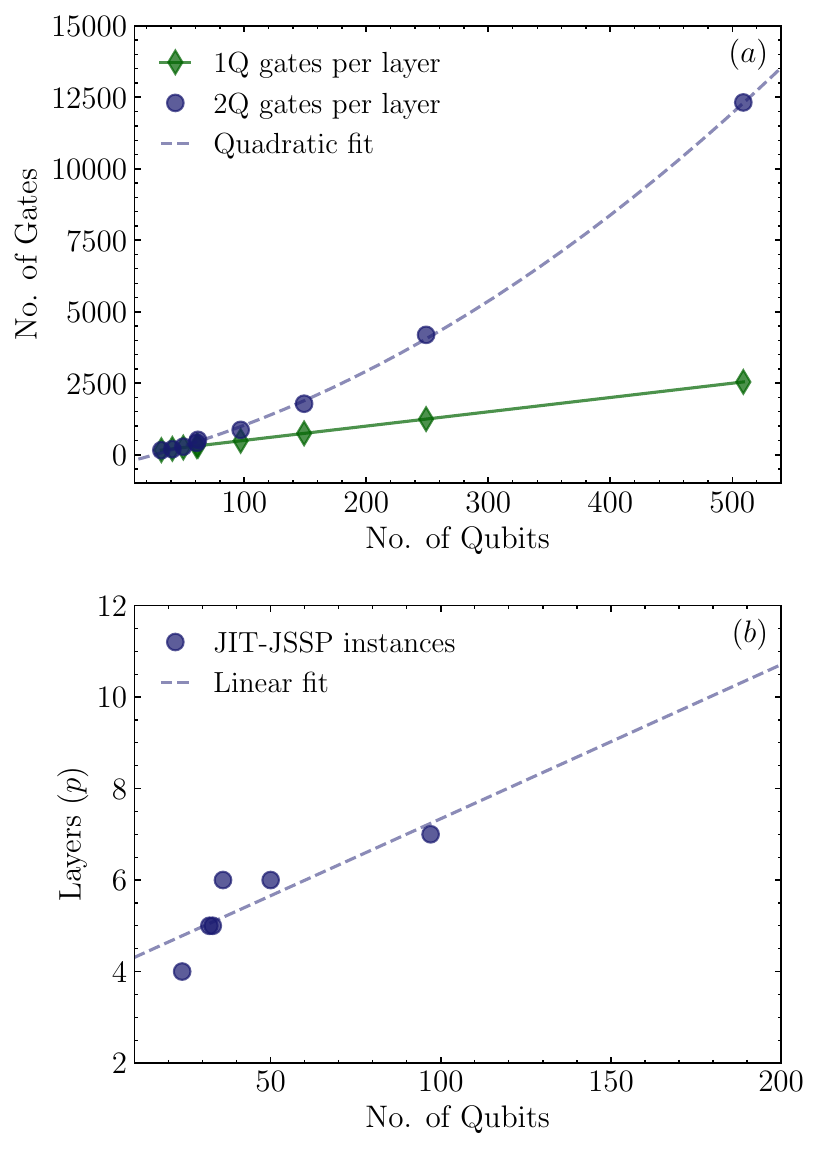}
    \caption{$(a)$ Scaling of single- and two-qubit gate counts per QAOA layer as a function of instance size (number of qubits). Data points correspond to JIT-JSSP sub-instances derived from the 1,300-qubit full instance. The three largest points represent circuits generated solely to construct the QAOA ansatz and extract gate counts (these circuits were not simulated or executed on quantm hardware). Two-qubit gates (native $R_{ZZ}$ on this class of hardware) follow an approximately quadratic scaling with instance size, as indicated by the dashed fit $y = 0.028x^2 + 10.493x - 298.214$. $(b)$ Number of QAOA layers $p$ used for each sub-instance executed on a simulator or quantum hardware in this work, representing the circuit depth required to achieve sufficient expressivity. A linear fit (dashed line), $y = 0.034x + 3.973$, provides an approximate extrapolation of the depth expected for larger instances of comparable expressivity.}
    \label{fig:gates}
\end{figure}

\noindent
The key results presented in the preceding sections are summarized in \cref{tab:results}. The classical computational time required to solve these instances scales approximately exponentially with the number of variables, as shown in \cref{fig:classical_comp_time} of \cref{sec:classical_solvers}.  Given that the number of variables scales at least quadratically with the number of jobs $J$ (specifically as $O(J^2M)$ when $T_m \approx J$), it follows that even moderate increases in problem size could potentially render the problem classically impractical. For instance, extrapolating the observed scaling suggests that a problem with $J=30$ jobs and $M=3$ machines (corresponding to at least 2,700 binary variables) would require approximately 4 days to solve using the benchmarked classical solver (see \cref{fig:classical_comp_time}).

Investigating the potential quantum scaling is therefore crucial. Our exploration using 50- and 97-qubit MPS simulations provides preliminary insights, albeit with limitations inherent to the tensor network approximation. Two key factors govern the feasibility of scaling quantum algorithms: qubit count and gate counts, where the latter is limited by gate fidelity, at least in the NISQ era. Extrapolating the observed scaling of required QAOA layers, as shown in \cref{fig:gates}(b), a 2,700-variable JIT-JSSP instance might require a $p \sim 10^2$ layer ansatz. Assuming quadratic scaling for the gate counts (see \cref{fig:gates}(a)), this corresponds to $\sim 10^7$ two-qubit gates, making it a promising target for fault-tolerant quantum computers. Additionally, ongoing improvements in gate fidelities, gate count pruning techniques, and the development of robust error mitigation methods \cite{Shaydulin2024-mf} can accelerate progress significantly. 

It is important to emphasize that Iterative-QAOA remains a heuristic algorithm and as such, the observed scaling behavior does not come with rigorous proofs. Nevertheless, exploring resource-efficient strategies, such as efficient qubit encodings \cite{Leidreiter2024-en, Schmid2024-ir, Bourreau2024-zd} or combining classical decomposition methods \cite{Ponce2023-ks,Pakhomchik2022-aw,El-Kholany2022-ks,Wu1999-bl} that break down large problem instances into manageable sub-problems, with quantum heuristics remains a promising direction for leveraging near-term quantum devices for complex optimization problems.

\section{Summary and Outlook}
\label{sec:outlook}
\noindent
In this work, we developed and evaluated a quantum heuristic method, Iterative-QAOA, designed to solve the NP-hard JSSP. This algorithm combines two resource-efficient strategies: a non-variational approach using shallow, fixed-parameter QAOA circuits and an iterative warm-starting process. The core of this method is an iterative mechanism that uses thermal averages of quantum measurement outcomes from one iteration to prepare a biased, more efficient initial state for the next iteration. This mechanism progressively guides the quantum search toward the low-energy sector of the cost Hamiltonian without increasing the depth of the quantum circuit.

Through numerical simulations and hardware executions, we benchmarked Iterative-QAOA against both its LR-QAOA baseline and the VarQITE algorithm. Our results confirm that a shallow-depth Iterative-QAOA significantly outperforms a deep-circuit LR-QAOA, converging to ground-state configurations with much higher probabilities. Compared to VarQITE, where we observed unfavorable scaling for this problem class, Iterative-QAOA demonstrated robust performance, finding high-quality solutions even for the 97-qubit instance. Furthermore, execution on trapped-ion quantum processors confirmed the algorithm's viability on near-term hardware where, even without significant error mitigation, it successfully identified optimal or high-quality sub-optimal solutions.

While the results presented are promising, we acknowledge that for the problem sizes studied, state-of-the-art classical heuristics can still achieve superior time-to-solution (see \cref{sec:classical_solvers}). The primary value of quantum heuristics such as Iterative-QAOA, however, lies in their potential for scalability to fault-tolerant quantum computers and their ability to explore solution spaces in fundamentally different ways. Although demonstrated here on the JSSP, the algorithm provides a general framework for combinatorial optimization. A key direction for future work is to apply this iterative method to other classes of problems, particularly those for which classical solution methods are known to scale poorly.

In addition, the Iterative-QAOA framework itself offers several avenues for refinement. In this work, we used a simple linear ramp schedule with the constraint $\Delta_\beta=\Delta_\gamma$, reducing the schedule to a single hyperparameter. Exploring non-linear schedules could lead to a more faithful approximation of the underlying adiabatic evolution, potentially yielding improved performance on larger problem instances. We also used a Boltzmann distribution to compute the weights of the linear combination of solutions that constructs the initial state of the algorithm. Exploring more sophisticated methods of constructing the initial state could also potentially accelerate the convergence of the algorithm leading to a better time-to-solution. Another promising direction is the use of more efficient qubit encodings \cite{Leidreiter2024-en, Schmid2024-ir, Bourreau2024-zd}. While such encodings could allow larger instances to be mapped on current hardware, this often comes at a cost of introducing higher-order local interactions in the Hamiltonian. Analyzing this trade-off, in the context of Iterative-QAOA, could be an interesting future research direction.

\begin{acknowledgments}
We would like to thank Felix Tripier, Yvette de Sereville, Andrew Arrasmith, and Andrii Maksymov for helpful discussions on DNL filtering and error mitigation techniques in general. This publication is supported by funding from The South Carolina Quantum Association.
\end{acknowledgments}

\bibliographystyle{quantum}
\bibliography{references}

\onecolumngrid
\appendix

\section{JSSP Instance Generation}
\label{sec:jssp-inst-gen}

\begin{table}[ht!]
    \centering
    \footnotesize
    \begin{tabularx}{0.6\columnwidth}{X c r}
    \toprule
        Parameter & Symbol & Value \\
        \midrule
         Machines & $M$ & 3 \\ 
         Jobs & $J$ & 20 \\
         Time Slots & $T_m$ & \{20, 22, 23\} \\
         Earliness Cost & $c_e$ & 1 \\
         Lateness Cost & $c_l$ & 3 \\
         Production Switching Cost & $c_p$ & 5 \\
         Penalty Weight & $\lambda$ & 10 \\
         \bottomrule
    \end{tabularx}
    \caption{Parameters for the full JIT-JSSP instance used in the experimental validation. The meaning of the variables are explained in~\cref{sec:probl_formulation}.}
\label{tab:exp_param}
\end{table}

\begin{figure*}[ht!]
    \centering
    \includegraphics[width=\textwidth]{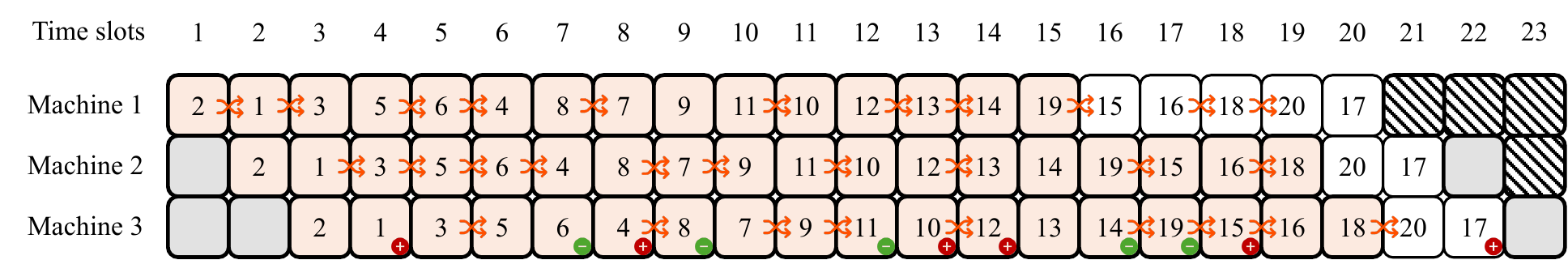}
    \caption{Gantt chart representing an optimal solution to the full instance and the time slots unfrozen in the 33-qubit instance. Each row corresponds to a machine and each column is a time slot that has an assigned job. Orange boxes represent fixed job assignments, gray boxes represent the fixed empty (idle) time slots assignments, while white boxes are free (unfrozen) time periods that are optimized. Dashed boxes denote empty time slots which were not classically optimized. Unit penalties are accrued for each production group switch (crossing arrows), early deliveries (green minus signs), and late deliveries (red plus signs). The production groups and due times of the underlying JIT-JSSP instance are defined in \cref{tab:prod_groups}.}
    \label{fig:gantt_33qubit}
\end{figure*}

\begin{table*}[ht!]
    \centering
    \footnotesize
    \begin{tabularx}{\textwidth}{lc *{20}{X}}
    \toprule
        & \multirow{2}{*}{Machine} & \multicolumn{20}{c}{Jobs}\\
        \cmidrule(lr){3-22} \\[-1em]
        & & 1 & 2 & 3 & 4 & 5 & 6 & 7 & 8 & 9 & 10 & 11 & 12 & 13 & 14 & 15 & 16 & 17 & 18 & 19 & 20 \\
        \cmidrule(lr){1-22} \\[-1em]
        \multirow{3}{*}{\makecell[l]{Production\\Groups}} & 1 & A & B & C & D & C & B & C & D & C & B & C & D & A & B & C & D & D & B & C & D \\
        & 2 & A & A & B & B & C & D & A & B & D & C & D & C & D & D & A & A & B & B & D & B \\
        & 3 & B & B & B & A & A & A & C & C & B & D & D & A & A & A & B & D & B & D & C & B \\
        \cmidrule(lr){1-22} \\[-1em]
        \makecell[l]{Due\\Times} & $-$ & 3 & 3 & 5 & 6 & 6 & 8 & 10 & 11 & 11 & 12 & 13 & 13 & 15 & 17 & 17 & 19 & 20 & 20 & 21 & 21\\
        \bottomrule
    \end{tabularx}
    \caption{Production groups and due times of the full JIT-JSSP instance.}
    \label{tab:prod_groups}
\end{table*}

\begin{table*}[ht!]
    \centering
    \footnotesize
    \begin{tabularx}{\textwidth}{Xlllllllll}
    \toprule
       \multirow{2}{*}{\makecell[l]{Free Variables \\ $(n_\text{var})$}} & \multicolumn{3}{c}{Machine 1} &  \multicolumn{3}{c}{Machine 2} &  \multicolumn{3}{c}{Machine 3}\\
        \cmidrule(lr){2-4} \cmidrule(lr){5-7} \cmidrule(lr){8-10} \\[-1em] 
        & Jobs & Time Slots & $n_1$ & Jobs & Time Slots & $n_2$ & Jobs & Time Slots & $n_3$ \\
        \midrule
        24 & 16-18, 20 & 17-20 & 4 & 17, 20 & 20-21 & 2 & 17, 20 & 21-22 & 2\\
        32* & 15-18, 20 & 16-20 & 5 & 17, 20 & 20-21 & 2 & 17, 20 & 21-22 & 2\\
        33 & 15-18, 20 & 16-20 & 5 & 17, 20 & 20-21 & 2 & 17, 20 & 21-22 & 2\\
        36 & 16-18, 20 & 17-20 & 4 & 16-18, 20 & 18-21 & 4 & 17, 20 & 21-22 & 2 \\
        50 & 15-18, 20 & 16-20 & 5 & 16-18, 20 & 18-21 & 4 & 17-18, 20 & 20-22 & 3\\
        97 & 15-20 & 15-20 & 6 & 15-20 & 16-21 & 6 & 15-18, 20 & 18-22 & 5\\
        \bottomrule
    \end{tabularx}
    \caption{Construction of JIT-JSSP sub-instances. Free (unfrozen) jobs and time slots used to generate all instances evaluated in this work. The number of free jobs/time slots at the end of the schedule for each machine ($n_1, n_2, n_3$) follows \cref{eq:nvar}, with the condition that $n_1 \ge n_2 \ge n_3$. The 32-variable instance (*) is derived from the 33-variable instance by additionally fixing variable $x_{1,15,20}$ to its optimal value of 0. The Gantt chart representing the 33-variable instance is shown in \cref{fig:gantt_33qubit}.}
    \label{tab:instances}
\end{table*}

\noindent
We designed a problem instance with $J=20$ jobs, each of which had to be processed on $M=3$ machines. On machine 1, all time slots were active (i.e., processed a job), and there were no idle slots. Following this, the number of time slots needed for machine 1 was $T_{1} = J= 20$. On machine 2, two time slots were allowed to be idle, corresponding to $T_{2}=J+2=22$, and on machine 3, three time slots were allowed to be idle, corresponding to $T_{3}=J+3 = 23$. 

The original formulation we solved is similar to the formulation presented in \cref{sec:probl_formulation}, with one key difference: it did not exogenously specify which time slots on each machine were idle. That is, it did not include sets of active ($A_m$) and idle ($I_m$) time slots for each machine $m$; rather, it followed the time assignment and idle slot constraints as presented in Ref. \citep{Amaro2022-or}. Solving the original instance produced the optimal solution presented in Fig. \ref{fig:gantt_33qubit}. After the original instance was generated, we used it to exogenously define the sets of active and idle time slots on each machine, essentially pre-defining whether a time slot on a machine would be active or idle. The resulting sets of idle time slots were $I_{1}=\emptyset, I_{2} = \{1,22\}, I_{3}=\{1,2,23\}$ (the grey boxes on Fig. \ref{fig:gantt_33qubit}) and active time slots on each machine $m$ were given by $A_m = \{1,\dots,T_m\} \setminus I_m$ (the orange and white boxes on Fig. \ref{fig:gantt_33qubit}). We used the resulting formulation with exogenously defined sets of active and idle slots (given in \cref{sec:probl_formulation}) for all analyses and discussion, and we refer to it throughout the paper as the ``full instance.'' Note that the optimal job-machine assignments in the original and full instances are the same.

We solved the full instance classically using Gurobi Optimizer software version 12.0.2 on an Intel(R) Core(TM) i7-8550U CPU @ 1.80GHz. This problem had 1,300 binary variables and 23 continuous variables; the continuous variables represent the due date and production cost penalties, far too many to be solved via a quantum simulation or current quantum hardware.
In order to end up with problems of suitable sizes for our simulator and hardware, we designed several sub-instances, where we strategically ``froze'' most variables in the full instance to their optimal value, but left a subset of the variables unfrozen, so that we could solve each sub-instance for these variables using quantum simulation and/or hardware. The free slots remain fixed once the full instance is solved, so each sub-instance only considers a subset of the $x_{mjt}$ variables. Ref.~\cite{Amaro2022-or} followed a similar process of freezing most variables, then solving for the unfrozen variables using quantum technology. Our goal was to generate a set of sub-instances that covered a wide range of number of variables, so that we could assess the performance of Iterative-QAOA across a range of problem sizes. The process of how we designed each sub-instance and determined the associated number of unfrozen variables is detailed below.

\begin{enumerate}
    \item We chose the number of slots (and therefore, the same number of jobs) to leave unfrozen on each machine, abiding by the following constraints:
    \begin{enumerate}
        \item We always freed up jobs at the end of each machine’s time slots.
        \item The number of free slots on machine 1 $\geq$ the number of free slots on machine 2 $\geq$ the number of free slots on machine 3.
    \end{enumerate}
    Note: These rules served two purposes. First, they prevented us from choosing trivial cases where visual inspection of the problem would allow it to be reduced to a smaller number of variables. Second, they ensured that we could use the same formula (shown in step 2) to calculate the number of variables in each sub-instance.
    
    \item We calculated the number of unfrozen variables associated with this set of free slots as follows
    \begin{align}
        n_1^2 + n_2^2 + n_3^2 = n_\text{var}.
        \label{eq:nvar}
    \end{align}
    This result is fairly intuitive, on machine $i$, we have $n_i$ free jobs, each of which can be placed in any of $n_i$ slots; therefore, we need to have an unfrozen variable for each potential job/slot pair, so we have $n_1 ^2$ variables for machine 1, $n_2^2$ variables for machine 2, and $n_3^2$ variables for machine 3.
    
    \item If the problem size calculated in step 2 was useful for us, we added it to the list of instances and returned to the first step to generate other instances as needed. If it was not useful for us, we returned to the first step to generate other instances. When our goal was to arrive at a specific number of variables, and this number couldn’t be achieved exactly via steps 1 and 2, we proceeded to step 4.
    
    \item Starting with the smallest sub-instance with more variables than our desired number, we refroze variables until we arrived at our desired number of variables (note that $n_\text{var}$ decreased by 1 for each variable that we refroze). We refroze variables whose optimal value was 0, because freezing variables whose optimal value is 1 would have resulted in a much less interesting problem. We now have our set of unfrozen variables, and this sub-instance is complete.
\end{enumerate}

In \cref{tab:exp_param}, we show all the parameter values used to construct the cost function (\cref{eq:jit-jssp_cost_func}) of the full problem instance, and \cref{tab:prod_groups} shows the production groups and due dates.

To illustrate the instance construction process, \cref{fig:gantt_33qubit} displays a specific sub-problem overlaid on an optimal solution of the full instance. In the Gantt chart, orange boxes represent variables fixed (frozen) to the their optimal values, while white boxes, denote the free (unfrozen) variables that define the sub-instance to be solved. Dashed boxes indicate empty time slots, arrows mark production group switches, and the plus and minus signs represent a late/early delivery of the job on machine 3, respectively. In this example, the number of free slots (white boxes) per machine are $n_1=5$, $n_2=2$, and $n_3=2$. According to \cref{eq:nvar}, this configuration corresponds to $n_\text{var}=33$ free binary variables. The 33-variable cost function (\cref{eq:jit-jssp_cost_func}) is then mapped onto a 33-qubit Hamiltonian (one-hot encoding) using the transformation \cref{eq:hamiltonian}. In \cref{tab:instances} we show the specific configurations of unfrozen jobs and time slots utilized to generate all sub-instances used in this work.

\section{Classical Solvers}
\label{sec:classical_solvers}

\begin{table}[t!]
    \centering
    \label{tab:classical_sol_2}
    \footnotesize
    \begin{tabularx}{0.6\columnwidth}{XXllll}
    \toprule
        \multirow{2}{*}{\makecell[l]{Time horizon\\ $[t,T_m]$}}  & \multirow{2}{*}{\makecell[l]{Free Variables\\$(n_\text{var})$}} & \multicolumn{4}{c}{Computational time [sec]}\\
        \cmidrule(lr){3-6} \\[-1em] 
        & & $\text{Run}\,1$ & $\text{Run}\,2$ & $\text{Run}\,3$ & Avg. \\[0.5em] 
        \midrule
        $[17,T_m]$ & 88 & 0.03 & 0.03 & 0.03 & 0.03\\
        $[16,T_m]$ & 123 & 0.04 & 0.05 & 0.03 & 0.04\\
        $[15,T_m]$ & 164 & 0.08 & 0.07 & 0.07 & 0.07 \\
        $[14,T_m]$ & 211 & 0.10 & 0.08 & 0.11 & 0.10\\
        $[13,T_m]$ & 264 & 0.21 & 0.12 & 0.11 & 0.15\\
        $[12,T_m]$ & 323 & 0.33 & 0.29 & 0.53 & 0.39\\
        $[11,T_m]$ & 388 & 0.51 & 0.48 & 0.46 & 0.48\\
        $[10,T_m]$ & 459 & 1.30 & 1.09 & 1.10 & 1.16\\
        $[9,T_m]$  & 536 & 2.14 & 2.03 & 1.96 & 2.04\\
        $[8,T_m]$  & 619 & 2.28 & 2.36 & 2.46 & 2.36\\
        $[7,T_m]$  & 708 & 3.73 & 3.59 & 3.35 & 3.56\\
        $[6,T_m]$  & 803 & 7.95 & 9.18 & 9.34 & 8.82\\
        $[5,T_m]$  & 904 & 28.25 & 32.83 & 32.56 & 31.21\\
        $[4,T_m]$  & 1011& 53.28 & 41.31 & 33.37 & 42.65\\
        $[3,T_m]$  & 1,124 & 112.62 & 98.76 & 131.12 & 114.17\\
        $[2,T_m]$  & 1,221 & 124.64 & 129.96 & 103.53 & 119.38\\
        $[1,T_m]$  & 1,300 & 226.00 & 229.74 & 229.93 & 228.56\\
        \bottomrule
    \end{tabularx}
    \caption{Time to solve to optimality on a classical computer for sub-instances of the full instance. The unfrozen (free) binary variables in each sub-instance are the binary decision variables between $t$ and $T_m$ (inclusive) that are optimized; this excludes any decision variable associated with a job $j$ assigned to machine $m$ in a time period prior to $t$.}
\end{table}

\begin{figure}[ht!]
    \centering
    \includegraphics[width=0.5\columnwidth]{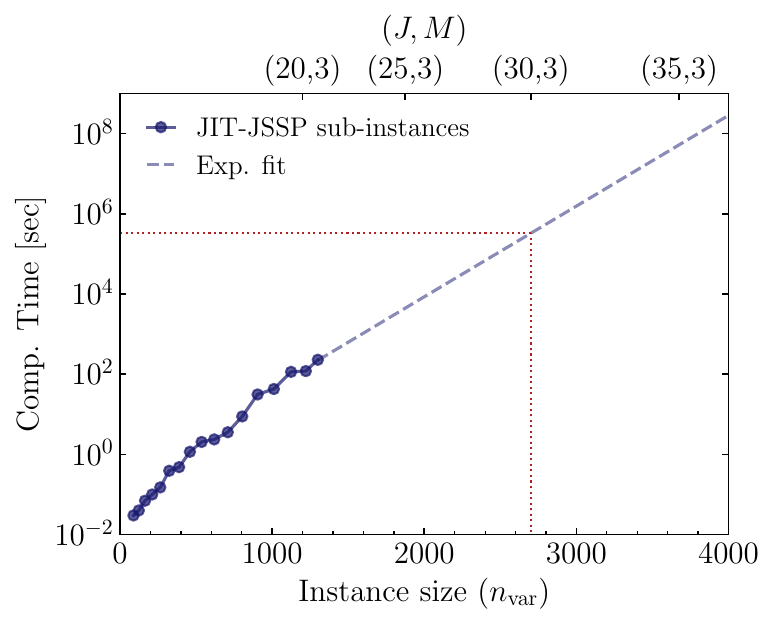}
    \caption{Classical computational time for sub-instances beginning at time period $t$. The data points correspond to the mean time-to-solution values from \cref{tab:classical_sol_2}. The dashed line shows an exponential fit, extrapolating the expected computational time for larger problems. The extrapolated computational time is only an approximate guide and will need to be verified by actual execution. The upper horizontal axis shows example combinations of jobs $J$ and machines $M$ that correspond to the lower bound of instance size $J^2M$. The dotted red line indicates a problem size for which the current formulation of the JSSP would take around 4 days to execute. This could represent a point beyond which quantum computation may begin to have advantage over classical computation. } 
    \label{fig:classical_comp_time}
\end{figure}

The JIT-JSSP instances we consider in this paper can also be solved using classical hardware. We present the time to solve to optimality using the Gurobi Optimizer (with presolve) software version 12.0.2 solved on a PC with Intel(R) Core(TM) i7-8550U CPU @ 1.80GHz. 
\cref{tab:classical_sol_2} shows the solution times to solve sub-instances with progressively fewer frozen variables. The time horizon $[t,T_m]$ associated with each row represents the time slots on machine $m$ that are optimized in a given sub-instance, i.e., with unfrozen variables; all other variables are frozen to the optimal solution presented in \cref{fig:gantt_33qubit}. Note that the row with the time horizon of [$1,T_m$] corresponds to the full instance. Exponential scaling in computational time as a function of instance size is shown in \cref{fig:classical_comp_time}.

\section{LR-QAOA Parameter Exploration}
\label{sec:lrqaoa_param_exploration}

\begin{figure*}[ht!]
    \centering
    \includegraphics[width=\textwidth]{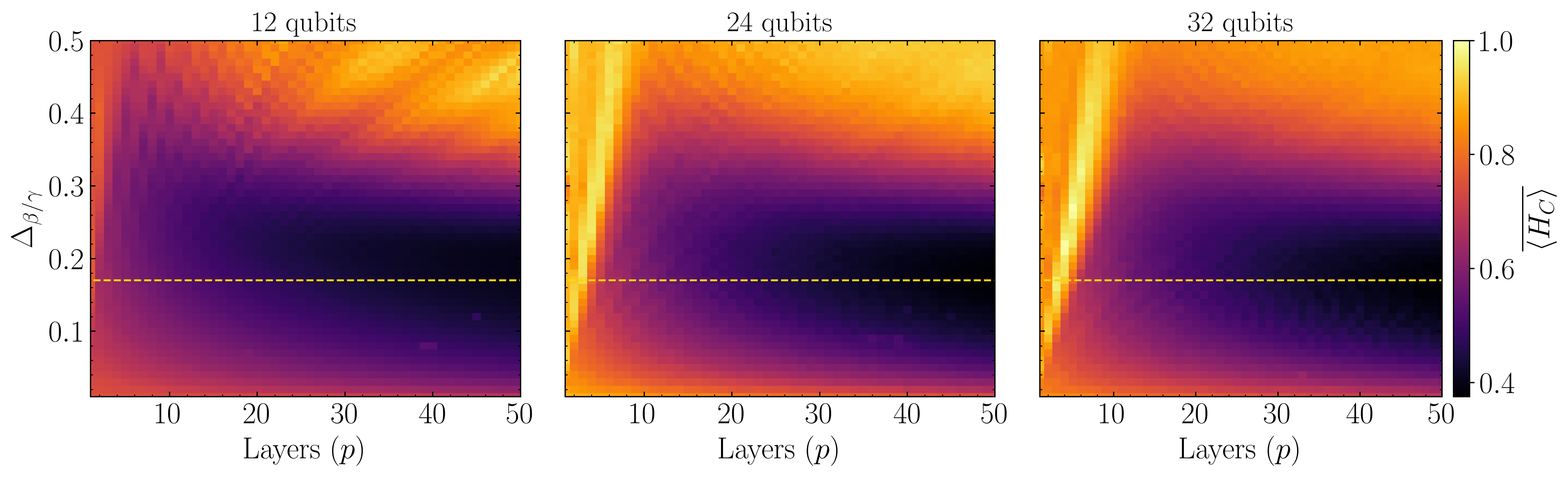}
    \caption{LR-QAOA performance landscape for JIT-JSSP instances of increasing size. Each panel correspond to a different problem instance: 12, 24, and 32 qubits (from left to right). The color scale shows the average cost $\langle H_C\rangle$ normalized by $\max\langle H_C\rangle$ as a function of QAOA layers $p$ and ramp parameter $\Delta_{\beta/\gamma}$. The dashed horizontal line at $\Delta_{\beta/\gamma}=0.17$ highlights the optimal parameter trajectory for all three instances.}
    \label{fig:lrqaoa_param_exp}
\end{figure*}

The performance of LR-QAOA is primarily determined by the ramp parameters $\Delta_\beta$ and $\Delta_\gamma$ and the number of layers $p$. To simplify the parameter search, we assume equal slopes, defining $\Delta_{\beta/\gamma}\equiv \Delta_\beta = \Delta_\gamma$. The optimal value of $\Delta_{\beta/\gamma}$ is identified by computing the expectation value of the cost Hamiltonian $\langle H_C\rangle$, from 4,000 measurement shots in ideal simulations for each point in the parameter landscape.

In \cref{fig:lrqaoa_param_exp} we present the landscapes for three JIT-JSSP instances of increasing size: 12, 24, and 32 qubits. For visual clarity, the color scale in each panel is normalized as
\begin{align}
    \frac{\langle H_C \rangle}{\max\langle H_C \rangle}, 
\end{align}
where $\max\langle H_C \rangle$ denotes the maximum expectation value computed across all values of $\Delta_{\beta/\gamma}$ and $p$ for a given problem instance; ensuring a uniform $z$-scale across all panels.

A key feature across all three cases is the emergence of a clear ``performance valley'' \cite{Kremenetski2023-vy, Montanez-Barrera2025-im}, a region where the algorithm most effectively minimizes the cost Hamiltonian as circuit depth increases. Notably, the optimal trajectory along this valley is consistently located at a ramp parameter of $\Delta_{\beta/\gamma}\approx0.17$, regardless of the problem size. We selected this value for all experiments in the main text, assuming this optimal parameter holds or deviates only slightly for larger instances. 

While the normalization $\langle H_C \rangle/\max\langle H_C \rangle$ helps visualize the ``performance valley'', the absolute performance of LR-QAOA is still size-dependent. The lowest average energy $\min\langle H_C\rangle$, increases as system size grows for a fixed $p$. This shows how the complexity of the problem increases with system size, therefore requiring deeper circuits to reach high-quality solutions. 

\section{Performance with sub-optimal LR-QAOA parameters}
\label{sec:sub-optimal_lrqaoa}
\noindent

\begin{figure*}[ht!]
    \centering
    \includegraphics[width=\textwidth]{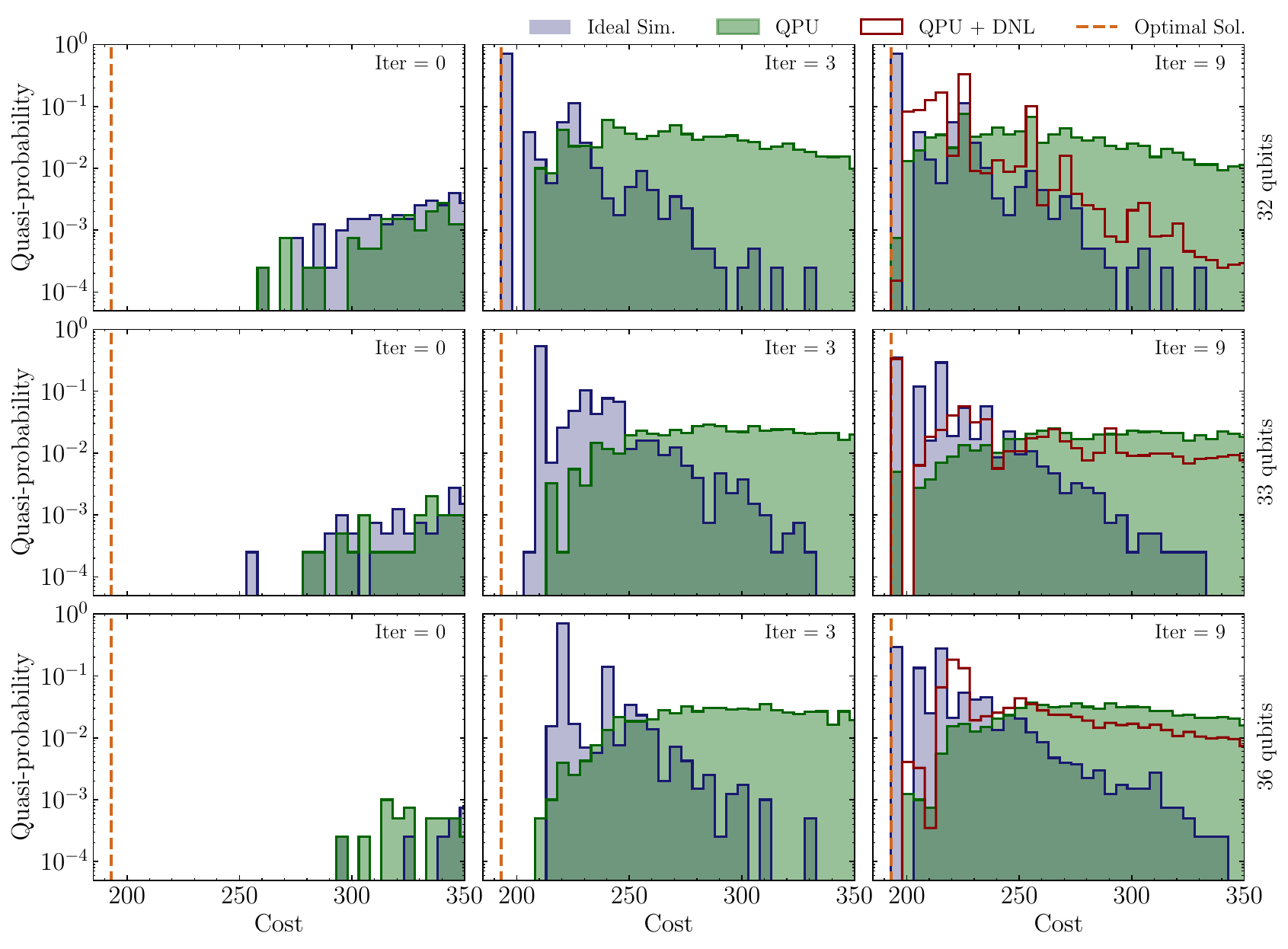}
    \caption{Performance of Iterative-QAOA on IonQ Forte Generation QPU on problems with 32, 33 and 36 qubits. These problems were executed with a sub-optimal LR-QAOA ramp parameter of $\Delta_{\beta/\gamma}=1.25$ and a fixed $\beta_T=0.5$. Even with sub-optimal parameters, the algorithm is robust enough to sample the optimal solution for the 32 qubits and 33 qubits problem instances and the first excited state in the 36 qubit problem instance when executed on quantum hardware.}
    \label{fig:sub-optimal_lrqaoa}
\end{figure*}

\begin{figure*}[t!]
    \centering
    \includegraphics[width=\textwidth]{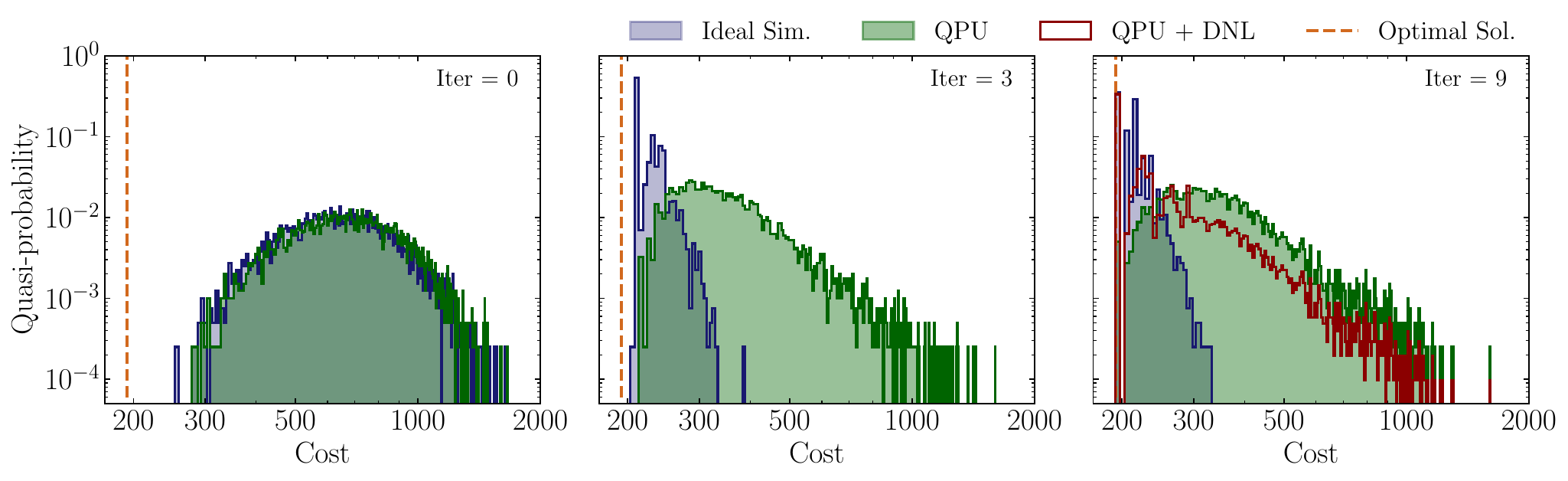}
    \caption{Performance of Iterative-QAOA on a 33-qubit JIT-JSSP instance executed on IonQ Forte generation QPU with a sub-optimal LR-QAOA ramp parameter of $\Delta_{\beta/\gamma}=1.25$ and a fixed $\beta_T=0.5$.}
    \label{fig:sub-optimal_lrqaoa_33q-full}
\end{figure*}

To study robustness, we evaluated the performance of Iterative-QAOA using a sub-optimal parameter schedule, selecting a ramp parameter of $\Delta_{\beta/\gamma}=1.25$; a value located far from the optimal performance valley (see \cref{fig:lrqaoa_param_exp}).

The results of these simulations and hardware runs for the same 32-, 33-, and 36-qubit instances are shown in \cref{fig:sub-optimal_lrqaoa,fig:sub-optimal_lrqaoa_33q-full}. Like in the main text, the two smallest instances were simulated using a statevector simulator, while the 36-qubit instance ideal results were obtained using an MPS simulator with a maximum bond dimension of $\chi=256$. Despite the sub-optimal schedule, the iterative process is still able to effectively drive the cost distributions towards the low-energy sector, successfully sampling the optimal solution in both ideal noiseless simulations and quantum hardware runs. While the final probabilities of sampling the ground state are lower than those achieved with the optimal schedule, this result underscores the robustness of the algorithm. It demonstrates that the iterative refinement of the initial state is a powerful mechanism that can compensate for a poorly chosen parameter schedule.

\section{Trapped-ion Quantum Hardware}
\label{sec:ion-traps-qpus}

\noindent
The experiments were performed on IonQ’s Forte-generation trapped-ion quantum processors, specifically the Forte and Forte Enterprise systems \cite{Chen2024-co}. Both devices employ chains of 36 $^{171}$Yb$^+$ ions, with quantum information encoded in the ground state hyperfine levels. Ions are produced via laser ablation and photoionization, and confined in a surface linear Paul trap integrated within a vacuum package.

Universal control is achieved through two-photon Raman transitions driven by $\SI{355}{nm}$ pulsed lasers, enabling arbitrary single-qubit rotations and native $R_{ZZ}$-type entangling gates. At the time of experimentation, direct randomized benchmarking (DRB) reported median two-qubit gate fidelities of 99.3\% (Forte) and 99.17\% (Forte Enterprise), with single-qubit fidelities near 99.98\%. Gate durations were approximately $\SI{130}{\micro s}$ for single-qubit and $\SI{950}{\micro s}$ for two-qubit operations.

A key architectural feature is the use of acousto-optic deflectors (AODs) for independent beam steering, which reduces crosstalk and alignment errors \cite{Kim:2008ApOpt,Pogorelov:2021PRXQ}. Combined with automated calibration and control protocols, this enables robust high-fidelity operation across extended ion chains.

\section{Error Mitigation: Debiasing with Non-Linear Filtering}
\label{sec:dnl}

\begin{figure*}[ht!]
    \centering
    \includegraphics[width=0.75\textwidth]{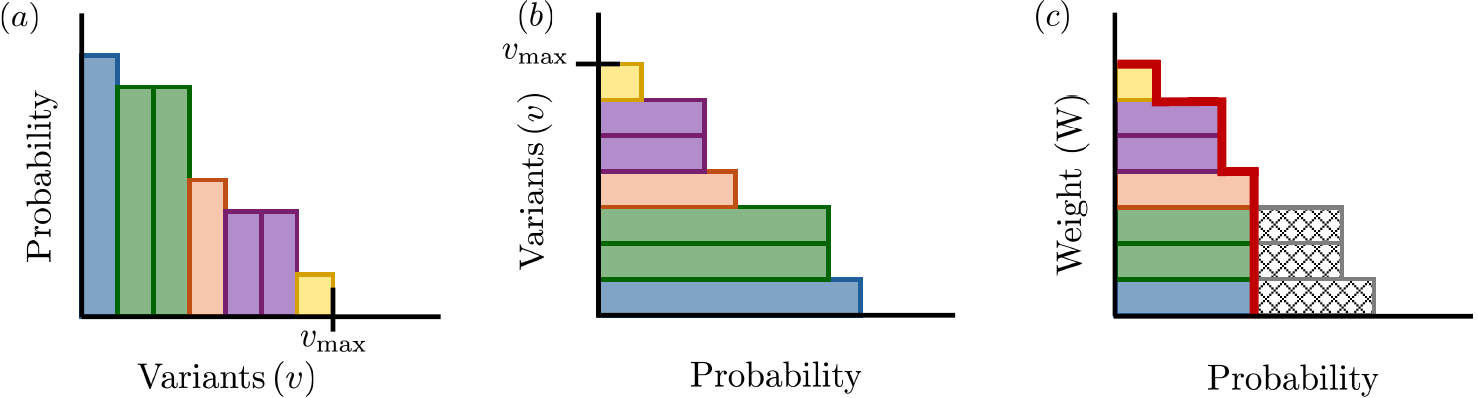}
    \caption{Aggregation process with non-linear filtering for a single measured bitstring \cite{Chernyshev2025-yd}. $(a)$ The distribution of the bitstring's probabilities is collected from each of the symmetrized circuit variants. $(b)$ These probabilities are then sorted and converted into a survival plot, which shows the number of variants that observed the bitstring at or above a given probability. The normalized area under this curve corresponds to the unmitigated probability, which is equivalent to a simple average of all variant outcomes. (c) Finally, a non-linear filter is applied by setting a threshold $v_\text{th}$, discarding bitstrings observed by fewer than a minimum number of variants. This step removes likely noise-induced artifacts that are inconsistent across the different circuit realizations, and the final, mitigated probability is calculated from the area under the filtered curve.
 }
    \label{fig:dnl}
\end{figure*}

\noindent
In this appendix we summarize the Debiasing with Non-linear Filtering (DNL) method used for error mitigation. DNL is a resource-efficient technique that suppresses hardware-induced errors by exploiting circuit and device symmetries. The method relies on executing multiple ``variants'' of a target circuit, which are equivalent in the noiseless limit but differ in their physical implementation. Variants can be generated, for example, by remapping logical qubits to different physical ions, using different gate decompositions, or by pairing circuits that differ only by final NOT gates. This construction symmetrizes dominant error sources such as location-dependent noise or measurement bias, enabling their suppression in post-processing.

The DNL aggregation algorithm goes as follows. For each measured bitstring, record and order its probability distribution across all variants (\cref{fig:dnl}a). Obtain the number of variants as a function of the measured probabilities of the bitstring and normalize the area under the curve (\cref{fig:dnl}b). Apply a non-linear weighting filter $W(v)$ (\cref{fig:dnl}c) which discards any bitstring observed by fewer than a minimum number of variants $v_\text{th}$. The specific weighting function used in this work is given by
\begin{align}
W(v) =\begin{cases}
\left(\dfrac{v}{v_{\max}}\right)^{\alpha}, & v > v_{\mathrm{th}}, \\[8pt]
0, & v \le v_{\mathrm{th}},
\end{cases}
\end{align}
where $v_\text{max}$ is the total number of variants and $\alpha>0$ controls how sharply outcomes observed in fewer variants are attenuated; we use $\alpha=4$. The final, mitigated probability for each bitstring is then calculated as a weighted average of its frequencies, with the weight for each bitstring determined by $W(v)$. This approach preserves bitstrings that are consistently observed across many variants while suppressing those that appear in only a few, which are treated as likely noise artifacts.

\end{document}